\documentstyle[preprint,aps,epsfig]{revtex}
\begin{document}
\draft
\preprint{
\vbox{
\halign{&##\hfil\cr
	& ANL-HEP-PR-98-27 \cr
	& JLAB-THY-98-11 \cr}}
}
\title{Massive Lepton Pairs as a Prompt Photon Surrogate}
\author{Edmond L. Berger$^a$, Lionel E. Gordon$^{b,c}$, and Michael Klasen$^a$}
\address{$^a$High Energy Physics Division,
             Argonne National Laboratory \\
             Argonne, Illinois 60439 \\
	$^b$Jefferson Laboratory, Newport News, VA 23606 \\
	$^c$Hampton University, Hampton, VA 23668}
\date{March 16, 1998}
\maketitle

\begin{abstract} 
We discuss the transverse momentum distribution for the 
production of massive lepton-pairs in hadron reactions at fixed target and 
collider energies within the context of next-to-leading order perturbative 
quantum chromodynamics.  For values of the transverse momentum $Q_T$ greater 
than roughly half the pair mass $Q$, $Q_T > Q/2$, we show that the 
differential cross section 
is dominated by subprocesses initiated by incident gluons.  Massive lepton-pair 
differential cross sections are an advantageous source of constraints 
on the gluon density, free from the experimental and theoretical complications 
of photon isolation that beset studies of prompt photon production.  We 
compare calculations with data and provide predictions for the differential 
cross section as a function of $Q_T$ in proton-antiproton reactions at 
center-of-mass energies of 1.8~TeV, and in proton-nucleon reactions at 
fixed target and LHC energies.  
\end{abstract} 
\vspace{0.2in}

\section{Introduction and Motivation}
\label{sec:1}

Both massive lepton-pair production, $h_1 + h_2 \rightarrow \gamma^* + X; 
\gamma^* \rightarrow l \bar{l}$, and prompt real photon production,  
$h_1 + h_2 \rightarrow \gamma + X$ are 
valuable probes of short-distance behavior in hadron reactions.  In 
addition to the opportunities they offer for tests of perturbative 
quantum chromodynamics (QCD), the two reactions supply critical information on 
parton momentum densities.  Massive lepton-pair production, commonly 
referred to as the Drell-Yan process~\cite{ref:DY}, provided early 
confirmation of three colors and of the size of next-to-leading contributions 
to the cross section differential in the pair mass Q.  The mass and 
longitudinal momentum (or rapidity) dependences of the 
cross section (integrated over the transverse momentum $Q_T$ of the pair) 
serve as laboratory for measurement of the {\it antiquark} momentum density, 
complementary to deep-inelastic lepton scattering from which one gains 
information of the sum of the quark and antiquark densities.  Inclusive prompt 
real photon production is a source of essential information on the 
{\it gluon} momentum density.  At lowest order in perturbation theory, the 
reaction is dominated at large values of the transverse momentum $p_T$ of the 
produced photon by the ``Compton" subprocess, $q + g \rightarrow \gamma + q$.  
This dominance is preserved at higher orders, indicating that the experimental 
inclusive cross section differential in $p_T$ may be used to determine the 
density of gluons in the initial 
hadrons~\cite{ref:BQ,ref:Baer,ref:Aurenche,ref:GV}.  

There are notable similarities in the theoretical analyses of the two
processes.  At first-order in the strong coupling strength, $\alpha_s$, 
the Compton subprocess and the annihilation subprocess $q +\bar{q} \rightarrow 
\gamma + g$ supply the transverse momentum of the {\it directly} produced 
prompt photons.  Identical subprocesses, with the real $\gamma$ replaced by a
virtual $\gamma^*$, are responsible at ${\cal O}(\alpha_s)$ for the transverse 
momentum of massive lepton-pairs.  An important distinction, however, is that 
fragmentation subprocesses play a very important role in prompt real photon 
production at colliders energies.  In these long-distance fragmentation 
subprocesses, the photon emerges from the fragmentation of a final parton, 
e.g., $q + g \rightarrow q + g$, followed by $q \rightarrow \gamma + X$.  
     
Another significant distinction is that interest in $h_1 + h_2 \rightarrow 
\gamma^* + X$ has been drawn most often to the domain in which the pair mass 
$Q$ is relatively large, justifying a perturbative treatment based on a small 
value of $\alpha_s(Q)$ and the neglect of inverse-power high-twist 
contributions (except near the edges of phase space).  On the other hand, 
interest in prompt real photon 
production is directed to the region of large values of $p_T$ with small 
$\alpha_s(p_T)$.  Interest in the transverse momentum $Q_T$ dependence of the 
massive lepton-pair production cross section has tended to be focussed on very 
small values of $Q_T$ where the cross section is largest.  Fixed-order 
perturbation theory~\cite{ref:Reno} is applicable for large $Q_T$, but it is 
inadequate at small $Q_T$, and all-orders resummation 
methods~\cite{ref:CSS,ref:DWS,ref:AEGM,ref:AK,ref:LY} 
have been developed to address the region $Q_T << Q$.  

In this paper we focus on the $Q_T$ distribution for $h_1 + h_2 \rightarrow 
\gamma^* + X$ in the region where $Q_T$ is greater than roughly half of 
the mass of the pair, $Q_T > Q/2$.  We present and discuss calculations 
carried out in next-to-leading order QCD at both fixed target and collider 
energies.  We show that the differential cross section in this region is 
dominated by subprocesses initiated by incident gluons.  Massive lepton-pair 
differential cross sections are therefore a valuable, heretofore perhaps 
overlooked, source of 
constraints on the gluon density.  We compare calculations with data and 
provide predictions for the differential cross section as a function of $Q_T$ 
in proton-antiproton reactions at a center-of-mass energy of 1.8~TeV, 
in proton-nucleon reactions at fixed-target momenta and proton-proton 
reactions at LHC energies.  

The requirement that both $Q$ be large and $Q_T > Q/2$ may well be punishing 
in terms of experimental rate, especially at fixed-target momenta.  However, 
as long as $Q_T$ is large, the perturbative requirement of small 
$\alpha_s(Q_T)$ can be satisfied without a large value of $Q$.  We therefore 
explore and advocate the potential advantages of studies of 
$d^2\sigma/dQ dQ_T$ as a function of $Q_T$ for modest values of $Q$, 
$Q \sim 2$GeV, below the range of the traditional Drell-Yan region.    
To be sure, one must 
avoid the region in $Q$ of prominent resonance formation, e.g., the regions of 
the $\rho$, $\rho'$, and $J/\psi$ resonances, among others.  In addition, 
there are various backgrounds with which to contend at small $Q$ such as the 
contributions to the event rate from prompt decays of heavy flavors, e.g., 
$h_1 + h_2 \rightarrow c + \bar{c} + X; c \rightarrow l + X$.  These heavy 
flavor contributions may be estimated by direct computation~\cite{ref:BerSop} 
and/or bounded through experimental measurement of the like-sign-lepton 
distributions.  Previous studies of the connection between massive lepton-pair 
production and prompt photon production are reported in 
Refs.~\cite{ref:halscott} and~\cite{ref:aurlow}.

As our calculations demonstrate, the $Q_T$ distribution of massive lepton 
pair production presents a valuable additional method for direct measurement of 
the gluon momentum distribution.  This method makes good use of a significant 
data base at collider and fixed target energies.  The method is similar in 
principle to the approach based on prompt photon production, but it avoids 
the experimental and theoretical complications of photon isolation that beset 
studies of prompt photon production.  

In $e^+e^-$ and in hadron-hadron reactions at collider energies,
prompt photons are observed and their cross sections are measured only
if the photons are relatively isolated in phase space.  Isolation is required  
to reduce various hadronic backgrounds including those from the 
electromagnetic decay of mesons, e.g., $\pi^o \rightarrow 2\gamma$.
The essence of isolation is that a cone of half-angle $\delta$ is drawn about 
the direction of the photon's momentum, and the isolated cross
section is defined for photons accompanied by less than a specified
amount of hadronic energy in the cone, e.g., $E_h^{\rm cone}\leq
E_{\rm max} = \epsilon_h E_{\gamma}$; $E_{\gamma}$ denotes the energy of the 
photon.  Instead of $\delta$, the Fermilab collider groups use the 
variable $R = \sqrt {(\Delta \phi)^2 + (\Delta \eta)^2}$, 
where $\Delta \eta$ and $\Delta \phi$ denote differences of rapidity and 
azimuthal angle variables.  Theoretical predictions will therefore depend 
upon the additional parameters $\epsilon_h$ and $\delta$ (or $R$).  The 
{\it isolated} cross section is not an {\it inclusive} cross section, and 
the usual factorization theorems for inclusive cross sections need not 
apply~\cite{ref:BGQ}.
Isolation removes backgrounds, but it also reduces the signal.  For example, 
it reduces the contribution from processes in which the photon emerges from 
the long-distance fragmentation of quarks and gluons, themselves produced in 
short-distance hard collisions.  
The necessity to invoke phenomenological fragmentation functions and the 
existing infrared ambiguity of the isolated cross section in next-to-leading 
order leaves some question about the extent to which isolated prompt photon 
data may be used for fully quantitative determinations of the gluon density.  
It is desirable to investigate other physical processes for extraction of the 
gluon density that are free from these systematic uncertainties.   

Fortunately, no isolation would seem necessary in the case of virtual photon 
production (and subsequent decay into a pair of muons) in typical collider or 
fixed target experiments.  Muons are observed only after they have penetrated a 
substantial hadron absorber.  Thus, any hadrons within a typical cone about 
the direction of the $\gamma^*$ will have been stopped, and the massive 
lepton-pair signal will be entirely inclusive.  For this reason, we claim 
that the cross section for massive lepton-pair production at large $Q_T$ is 
an {\it advantageous} source of information on the gluon density.  
   
In Sec.~II, we review the formalism of the next-to-leading order calculations 
of the transverse momentum distribution for both prompt photon production and 
massive lepton-pair production, demonstrating the analytic similarities and 
differences.  In Sec.~III, we present numerical calculations that establish 
the regions of $Q_T$ in which the gluon initiated Compton subprocess 
dominates the differential cross section $d^2\sigma/dQ dQ_T$.  Comparisons with 
data and predictions are provided in Sec.~IV.  Fixed-target 
data~\cite{ref:expt1} are limited to the region $Q_T < Q$, but they are useful 
for establishing the rough value in $Q_T \simeq Q/2$ above which resummation 
may be ignored.  Collider data~\cite{ref:expt2} extend to larger values of 
$Q_T$.   Our conclusions are summarized in Sec.~V.  


\section{Massive Lepton Pair Production and Prompt Photon Production at 
Next-to-leading Order}
\label{sec:2}

In inclusive hadron interactions at collider energies, $h_1 + h_2 \rightarrow
\gamma^* + X$ with $\gamma^* \rightarrow l \bar{l}$, lepton pair
production proceeds through partonic hard-scattering processes involving
initial-state light quarks $q$ and gluons $g$. In lowest-order QCD, at
${\cal O}(\alpha_s^0)$, the only partonic subprocess is $q + \bar{q}
\rightarrow \gamma^*$. At ${\cal O}(\alpha_s)$, both $q + \bar{q} \rightarrow
\gamma^* + g$ and $q + g \rightarrow \gamma^* + q$ participate, with the
recoil of the final parton balancing the transverse momentum of the
lepton-pair. These processes are shown in Figs.~1(a) and 2(a). Calculations
of the cross section at order ${\cal O}(\alpha_s^2)$ involve virtual gluon loop
corrections to these ${\cal O}(\alpha_s)$ subprocesses (Figs.~1(b) and 2(b))
as well as real gluon radiation contributions from a wide range of 
$2 \rightarrow 3$ parton subprocesses (of which some examples are shown in 
Figs.~1(c) and 2(c)).

The physical cross section is obtained through the factorization theorem,
\begin{equation}
 \frac{d^2\sigma_{h_1h_2}^{\gamma^*}}{dQ_T^2dy} = \sum_{ij} \int dx_1 dx_2
 f^i_{h_1}(x_1,\mu_f^2) f^j_{h_2}(x_2,\mu_f^2) 
 \frac{sd^2\hat{\sigma}_{ij}^{\gamma^*}}
 {dtdu}(s,Q,Q_T,y;\mu_f^2).
 \label{dy1}
\end{equation}
It depends on the hadronic center-of-mass energy $S$ and on the mass $Q$, the
transverse momentum $Q_T$, and the rapidity $y$ of the virtual photon; $\mu_f$ 
is the factorization scale of the scattering process.  The
usual Mandelstam invariants in the partonic system are defined by $s =
(p_1+p_2)^2,~t = (p_1-p_{\gamma^*})^2$, and $u = (p_2-p_{\gamma^*})^2$,
where $p_1$ and $p_2$ are the momenta of the initial state partons and
$p_{\gamma^*}$ is the momentum of the virtual photon.  The indices $ij \in
\{q\bar{q},qg\}$ denote the initial parton channels whose contributions are 
added incoherently to yield the total physical cross section. The partonic cross
section $\hat\sigma_{ij}^{\gamma^*}(s,Q,Q_T,y;\mu_f^2)$ is obtained commonly
from fixed-order QCD calculations through
\begin{equation}
 \frac{d^2\hat{\sigma}_{ij}^{\gamma^*}}{dtdu} =
   \alpha_s  (\mu^2) \frac{d^2\hat{\sigma}_{ij}^{\gamma^*,(a)}}{dtdu}
 + \alpha_s^2(\mu^2) \frac{d^2\hat{\sigma}_{ij}^{\gamma^*,(b)}}{dtdu}
 + \alpha_s^2(\mu^2) \frac{d^2\hat{\sigma}_{ij}^{\gamma^*,(c)}}{dtdu}
 + {\cal O} (\alpha_s^3).
 \label{dy2}
\end{equation}
The tree, virtual loop, and real emission contributions are labeled
(a), (b), and (c) as are the corresponding diagrams in Figs.~1 and 2.  The 
parameter $\mu$ is the renormalization scale.  It is set equal to the 
factorization scale $\mu_f = \sqrt{Q^2+Q_T^2}$ throughout this paper.

The cross section for $h_1 + h_2 \rightarrow \l\bar{l} + X$, differential
in the invariant mass of the lepton pair $Q^2$ as well as its transverse 
momentum and rapidity, is obtained from Eq.~(\ref{dy1}) by the relation
\begin{equation}
 \frac{d^3\sigma_{h_1h_2}^{l\bar{l}}}{dQ^2dQ_T^2dy} = \left(
 \frac{\alpha_{em}}{3\pi Q^2} \right) \frac{d^2\sigma_{h_1h_2}^{\gamma^*}}
 {dQ_T^2dy}(S,Q,Q_T,y), 
 \label{dy3}
\end{equation}
where $Q^2 = (p_l + p_{\bar l})^2$, and $p_l, p_{\bar l}$ are the four-momenta 
of the two final leptons. 
The so-called Drell-Yan factor $\alpha_{em}/(3\pi Q^2)$ is included in all
the numerical results for virtual photon production presented in
Secs.~III and IV.

As stated in Sec.~I, the partonic subprocesses that contribute to lepton pair 
production and to prompt photon production at finite transverse momentum 
are identical except for the fact that the photon is off-shell in the case of 
massive lepton pair production.  As expected, the ${\cal O}(\alpha_s)$ cross 
section formul{\ae} for virtual photon production reduce to those for real 
photon production in the limit $Q^2 \rightarrow 0$ \cite{ref:halscott}.  It 
was also noted earlier that fragmentation processes, which are not present in 
the Drell-Yan case, are important for real photon production.  Taking into 
account direct and fragmentation pieces, we express the inclusive cross 
section for prompt photon production $h_1 + h_2 \rightarrow \gamma + X$ as 
\begin{eqnarray}
 \frac{d^2\sigma_{h_1h_2}^{\gamma}}{dp_T^2dy} &=& \sum_{ij} \int dx_1 dx_2
 f^i_{h_1}(x_1,\mu_f^2) f^j_{h_2}(x_2,\mu_f^2) \label{dy4}\\
 && \left[ \frac{sd\hat{\sigma}_{ij}^{\gamma,~direct}}{dtdu}(s,p_T,y;\mu_f^2,
\mu_F^2)
 + \sum_k\int\frac{dz}{z^2}D_k^{\gamma}(z,\mu_F^2)
 \frac{sd\hat{\sigma}_{ij}^{k,~fragm}}{dtdu}(s,p_T,y;\mu_f^2,\mu_F^2) \right].
 \nonumber
\end{eqnarray}
The transverse momentum of the real photon is $p_T$. 
The fragmentation function $D^{\gamma}_k(z,\mu_F^2)$ is the probability for 
emission of a photon with momentum fraction $z$ from parton $k$ at the 
fragmentation scale $\mu_F^2$.  The function is a non-perturbative quantity 
that cannot be predicted by QCD but instead requires input from experiment at 
some reference scale. Its evolution to greater values of the scale is 
specified by perturbation theory. In this paper, we set $\mu_F = \mu_f = \mu$.  

In prompt photon production, fragmentation contributes at leading order
${\cal O}(\alpha_s)$.  Although the hard subprocess matrix elements 
$\hat{\sigma}_{ij}^{k,~fragm}$ in the 
fragmentation case are ${\cal O}(\alpha_s^2)$, the fragmentation function has 
a logarithmic dependence on the scale $\mu_F^2$ and is effectively of 
${\cal O}(1/\alpha_s)$.  When the subprocess matrix elements are convoluted 
with the fragmentation function, the product is of ${\cal O} (\alpha_s)$.

At next-to-leading order ${\cal O} (\alpha_s^2)$, one must calculate virtual
loop corrections to the leading order processes (Figs.~1(b) and 2(b)) and the
three-body processes $q+g\rightarrow\gamma^*+q+g$, $q+\bar{q}\rightarrow
\gamma^*+g+g$, $q+q\rightarrow \gamma^*+q+q$, $g+g \rightarrow \gamma^*+q
+\bar{q}$, $q+\bar{q}\rightarrow \gamma^*+q+\bar{q}$, $q+q'\rightarrow\gamma^*
+q'+q$, $q+\bar{q}'\rightarrow \gamma^*+q'+\bar{q}'$, and $q+\bar{q}
\rightarrow\gamma^*+q'+\bar{q}'$ (Figs.~1(c) and 2(c)). The $2 \rightarrow 3$
parton 
matrix elements are integrated in $4-2\epsilon$ dimensions over the singular
regions of phase space to expose the infrared and collinear singularities as
$1/\epsilon$-poles. These poles are then canceled against the poles present in 
the virtual corrections. In general, this procedure applies to virtual photons
as well as real photons.  If the limit $Q^2 \rightarrow 0$ is taken carefully, 
and no overall powers of $Q^2$ are factored out, the results for the virtual 
photon cross section will again reduce to those for the real photon cross 
section, as they do in leading order.

There is, however, one particular case that has to be treated separately.
In a next-to-leading order calculation of prompt photon production, a 
singularity is produced whenever the photon becomes collinear to a final state 
quark in the integration of the matrix elements over phase-space. This
collinear singularity must be subtracted and absorbed into the fragmentation
function for a quark to produce a photon. The subtraction is done at an 
arbitrary 
fragmentation scale $\mu_F$, and hence the direct photon cross section acquires
a logarithmic dependence on the scale $\mu_F$ that is partially canceled
when fragmentation contributions are included. Fragmentation contributions
are not present in the virtual photon case since the photon is 
off-shell, and its invariant mass $Q$ regulates the singularity. This finite 
mass regularization leads to 
a logarithmic dependence of the Drell-Yan cross section on the virtuality of
the photon or the invariant mass of the observed lepton-pair $Q$.
As long as $Q$ is large enough, these logarithms are small so that
no resummation/factorization is necessary for a reliable prediction of
the cross section. The Drell-Yan cross section will not have a
fragmentation contribution except in the limit of very small $Q$.

As an example of the discussion in the paragraph above, we consider the 
subprocess $q\bar{q}\rightarrow q'\bar{q}'\gamma(\gamma^*)$ for prompt photon 
production and for the Drell-Yan case. The relevant diagrams are shown
in Fig.~3(a) and 3(b) for initial- and final-state photon radiation. Since 
we are interested mainly in the structure of the final state
in the region where a photon becomes collinear to a quark, this subprocess is
typical and any conclusions drawn from it can be applied to the more
complicated processes such as $qg$ scattering.  In this discussion
we concentrate on the final-state collinear regions, but we use the
full three-body matrix elements for both Drell-Yan~\cite{ref:CCFG} and prompt
photons~\cite{ref:GV} integrated fully over phase space, before factorization of
any singularities. The exact forms for the phase space integrals can be found in
Ref.~\cite{ref:GV} for prompt photons and in Refs.~\cite{ref:EMP,ref:CCFG} for 
the Drell-Yan case.
For the direct photon cross section $s d^2\hat{\sigma}_{ij}^{\gamma,~direct}
/dtdu$, the phase space integrals yield a singular term of the form 
\begin{equation}
\frac{1}{\epsilon}2 K_{\gamma}\frac{-(t^2 + u^2)(2 s^2 + 2 s t + t^2 + 2 s u
+ 2 t u + u^2)}
  {s (t + u)^4} = \frac{1}{\epsilon}f(s,t,u),
\label{dy5}
\end{equation}
whereas the corresponding logarithmic term in the Drell-Yan case is
\begin{eqnarray}
 && \ln\left[ \frac{s+Q^2-s_2+\lambda}{s+Q^2-s_2-\lambda}\right] K_{DY} \\
 && \left[ \hspace*{0.2cm}
 \frac{1}{\lambda^5}\left( \frac{3 Q^2 u^2(u - t)(t + u -2 s_2)}{s} 
 + \frac{3 (s - Q^2) u^2 (u^2 - t^2)}{Q^2 + s - s_2} \right) \right. \nonumber \\
 && + \left. \frac{1}{\lambda^3}\left(
  \frac{u (2 s s_2 - 2 s_2^2 + 2 s_2 t - s u + 4 s_2 u - t u - 3 u^2)}{s} 
  \right. \right. \nonumber \\
 && \left. \left. \hspace*{1cm}+ \frac{u^2}{Q^2 + s - s_2}\left(-2s-t+\frac{t^2}{s}+3u-
   \frac{u^2}{s}\right) \right) \right. \nonumber \\
 && + \left.
  \frac{1}{\lambda}\left( \frac{1}{s}\left( \frac{3 s}{4} - \frac{s_2}{2}
  + u \right) +
     \frac{1}{Q^2 + s - s_2}\left( \frac{3 s}{2} + \frac{5 u}{2} +
     \frac{2 u^2}{s} \right) \right) \right] \nonumber\\
 && + \hspace*{0.2cm}(t \longleftrightarrow u) \nonumber \\
 && = \ln\left[ \frac{s+Q^2-s_2+\lambda}{s+Q^2-s_2-\lambda}\right]
 g(s,t,u,Q^2). \nonumber
\end{eqnarray}
The quantities $K_{\gamma}$ and $K_{DY}$ represent products of color, electric 
charge, and phase space factors.  The function 
$\lambda = \sqrt{(t+u)^2-4 Q^2 s_2}$; $s$, $t$, and $u$ are defined just 
below Eq.~(\ref{dy1}), and $s_2 = s + t + u - Q^2$.

The functions $f(s,t,u)$ and $g(s,t,u,Q^2)$ differ, of course, due to the
$Q^2$-dependence in the Drell-Yan case. In the limit $Q^2\rightarrow 0$, 
however, $g(s,t,u,Q^2)\rightarrow f(s,t,u)$, as expected. The logarithm 
in Eq.~(6) becomes infinite in this limit and thus corresponds to the
pole $1/\epsilon$.  For the prompt photon case the pole term is subtracted
in some factorization scheme by the addition of a counter term. In the
$\overline{\rm MS}$ scheme the counter term is schematically 
\begin{equation}
K(\epsilon)\frac{1}{\epsilon}\left( \frac{\mu^2}{\mu_F^2}\right)^{\epsilon}
\hat{\sigma}^{q\bar{q}\rightarrow q' \bar{q}'}P_{\gamma/q}(z),
\end{equation}
where $\hat{\sigma}^{\bar{q}q}$ is the hard $2$-body subprocess matrix
element, and $P_{\gamma/q}(z)$ is the usual splitting function to produce
a photon from a quark. This procedure replaces the pole term with new
terms depending on $\ln(\mu_F^2)$.

To summarize this Section, we have shown that the cross section formul{\ae}
for virtual and real photon production in hadronic collisions are intimately
related in leading order ${\cal O}(\alpha_s)$ as well as in next-to-leading
order ${\cal O}(\alpha_s^2)$. If the logarithmic terms in $Q^2$ are first
isolated, setting $Q^2 = 0$ in the rest of the Drell-Yan cross section leads
to a recovery of the prompt photon cross section except for the $\ln(\mu_F^2)$
terms and numerically small finite terms associated with the factorization
of the collinear photon singularity into the fragmentation function.
The situation is completely analogous to hard photoproduction where the
photon participates in the scattering in the initial state instead of
the final state. For real photons, one encounters an initial-state
singularity that is factorized into a photon structure function. For
virtual photons, this singularity is replaced by a logarithmic
dependence on the photon virtuality $Q$ \cite{ref:KKP}.


\section{Dominance of the Compton Subprocesses for $Q_T > Q/2$}
\label{sec:3}

Having identified in Sec.~II the subprocesses that contribute in 
next-to-leading order to 
massive lepton-pair production, $h_1 + h_2 \rightarrow \gamma^* + X$, 
and to prompt real photon production, $h_1 + h_2 \rightarrow \gamma + X$,  
and having established the correspondence between the analytic forms of the 
contributions to these two reactions, we turn in this Section to explicit 
evaluations of the differential cross sections as functions of $Q_T$ and 
$p_T$, respectively, at both fixed-target momenta and collider energies.  
We work in the $\overline{\rm MS}$ renormalization scheme and set the 
renormalization and factorization scales equal.  We employ 
the CTEQ4M set of parton densities~\cite{ref:CTEQ} and a two-loop 
expression for the strong coupling strength $\alpha_s(\mu)$, with five 
flavors and appropriate threshold behavior at $\mu = m_b$.  For the massive 
lepton-pair case, $\alpha_s$ is evaluated at a hard scale 
$\mu = \sqrt{Q^2+Q_T^2}$, and for prompt real photon production, $\mu = p_T$.  

Beginning with $p + N \rightarrow \gamma^* + X$ at $p_{lab} = 800$ GeV, we 
present the invariant inclusive cross section $Ed^3\sigma/d p^3$ as a function 
of $Q_T$ in Fig.~4.  Shown in this figure are the $q {\bar q}$ and $q g$ 
perturbative contributions 
to the cross section at leading order and at next-to-leading order.  For 
this calculation, we integrate the invariant inclusive cross section over the 
scaled longitudinal momentum range 0.1 $< x_F<$ 0.3 and over the mass interval 
5 $<Q<$ 6 GeV, and we divide by the bin width in $x_F$.  To represent the 
target nucleon N, we use parton densities equal to 50 \% of the proton 
densities and 50 \% of the neutron densities, having in mind a comparison with 
deuterium data from Fermilab experiment E772, to be discussed in Sec. IV.  

The results in Fig.~4 show that the leading-order and the next-to-leading 
order contributions are about equal in size for $Q_T >$ 1~GeV in both the 
$q {\bar q}$ and $q g$ channels, meaning that the cross section through 
next-to-leading order is about twice that of the leading-order value.  For 
$Q_T<$ 2 GeV, the $q {\bar q}$ contribution exceeds that of $q g$ channel. 
However, for values of $Q_T >$ 2 GeV, the $q g$ contribution becomes 
increasingly important.  As shown in Fig.~5(a), the $q g$ contribution 
accounts for about 80 \% of the rate once $Q_T \simeq Q$.  The results in 
Fig.~5(a) also demonstrate that subprocesses other than those initiated by the 
$q {\bar q}$ and $q g$ initial channels are of negligible import.

In Fig.~5(b), we display the fractional contributions to the cross section 
as a function of $Q_T$ for a larger value of Q: 11 $<Q<$ 12 GeV.  In this 
case, the fraction of the rate attributable to $qg$ initiated subprocesses 
again increases with $Q_T$. It becomes 80 \% or greater for $Q_T \simeq Q/2$.  

For the calculations reported in Figs.~4 and~5(a,b), we chose values of Q in 
the traditional range for studies of massive lepton-pair production, viz., 
above the interval of the $J/\psi$ and $\psi'$ states and either below 
or above the interval of the $\Upsilon's$.  

For Fig.~5(c), we select the interval 2.0 $<Q<$ 3.0 GeV.  In this region, one 
would be inclined to doubt the reliability of leading-twist perturbative 
descriptions of the cross section $d\sigma/dQ$, {\it integrated} over all 
$Q_T$.  However for values of $Q_T$ that are large enough, a perturbative 
description of the $Q_T$ dependence of $d^2\sigma/dQdQ_T$ ought to be 
justified.  The results presented in Fig.~5(c) demonstrate that, as at 
higher masses, the $qg$ incident subprocesses dominate the cross section 
for $Q_T$ larger than a few GeV.  
 
We return to $pN$ reactions at fixed-target momenta in Sec.~IV and discuss 
comparisons of our cross sections with data, but we address first the 
case of $p +  \bar {p} \rightarrow \gamma^* + X$ at collider energies.  

In Fig.~6, we display the theoretical invariant inclusive cross section at 
the energy of the FNAL collider energy as a function of $Q_T$ for four regions 
of $Q$.  In addition to the regions of $Q$ mentioned above, we select a region 
of relatively large $Q$, 30.0 $< Q <$ 35.0 GeV, to extend the region of 
interest of our analysis.  For these results at 
collider energies, we average over the rapidity interval -1.0 $<y<$ 1.0.  
The fractional contributions from the $qg$ and $q {\bar q}$ subprocesses up 
through next-to-leading order are 
shown in Fig.~7.  Evident in Fig.~7 is that the $qg$ subprocess is the most 
important subprocess as long as $Q_T > Q/2$.  The dominance of the $qg$ 
subprocess diminishes somewhat with $Q$, dropping from over 80 \% for the 
lowest values of $Q$ to about 70 \% 
at its maximum for $Q \simeq$ 30 GeV.  In addition, for very large $Q_T$, the 
significant luminosity associated with the valence dominated ${\bar q}$ 
density in $p {\bar p}$ reactions begins to raise the 
fraction of the cross section attributed to the $q {\bar q}$ subprocesses.    

The calculations presented in Figs.~4 - 7 show convincingly that data on the 
transverse momentum dependence of the cross section for massive lepton-pair 
production at both fixed-target and collider energies should be a very 
valuable independent source of information on the gluon density.  It is 
instructive to compare the results at collider energies with those expected 
for prompt real photon production.  In Fig.~8, we present the predicted 
differential cross section for prompt photon production in the extreme 
cases: (a) all contributions, with full fragmentation taken into 
consideration (solid line), and (b) no fragmentation contributions included 
(dashed line).  The breakdown of these two cross sections into their 
$q {\bar q}$ and $qg$ components is presented in Fig.~9.  Comparing the 
magnitudes of the prompt photon and massive lepton pair production cross 
sections in Figs.~6 and~8, we note that the inclusive prompt photon cross 
section is about a factor of 400 greater than the massive lepton-pair cross 
section integrated over the mass interval 2.0 $< Q <$ 3.0 GeV.  This factor is 
attributable in large measure to the factor $\alpha_{em}/(3 \pi Q^2)$ 
associated 
with the decay of the virtual photon to $\mu^+ \mu^- $.  At current integrated 
luminosities, prompt photons have been observed with values of $p_T$ extending 
to 100 GeV and beyond~\cite{ref:cdfdat}.  It appears that it should be 
possible to examine massive lepton-pair cross sections in the same data sample 
out to $Q_T$ of 30~GeV or more.  As may be appreciated from a comparison of 
Figs.~7 and~9, dominance of the $qg$ contribution in the massive lepton-pair 
case is as strong if not stronger than it is in the prompt photon case.  
The statistical limitation to $Q_T$ of 30 GeV or so in the current data 
sample means that the reach in $x_{gluon}$, the fractional light-cone 
momentum carried by the incident gluon, is limited presently to 
$2Q_T/\sqrt S \sim 0.033$ or so, about a factor of three less than that 
potentially accessible with prompt photons in the same sample of data.    
It is valuable to be able to investigate the gluon density in the region 
$x_{gluon} \sim 0.033$, and less, with a process that has reduced 
experimental and theoretical systematic uncertainties from those of the 
prompt photon case.  


\section{Comparisons with Data and Predictions}
\label{sec:4}
In this Section we compare our calculations with data 
on massive lepton-pair production at large values of $Q_T$, try to establish 
the domain in $Q_T$ over which the next-to-leading order perturbative 
calculations should be reliable, and offer predictions in addition to 
those presented in Sec.~III.  

In Fig.~10, we show the invariant inclusive cross section $Ed^3\sigma/d p^3$ as 
a function of $Q_T$ for $p + {\bar p} \rightarrow \gamma^* +X$, with 
$\gamma^* \rightarrow \mu^+ \mu^-$, at $\sqrt S =$ 630 GeV, with 
$2m_{\mu} < Q <$ 2.5 GeV.  These results are averaged over the rapidity 
interval -1.7 $< y <$ 1.7.  We present the next-to-leading order perturbative 
cross section along with its two major components, the $qg$ and $q {\bar q}$ 
contributions.  The theoretical expectation is in good agreement with the data 
published by the CERN UA1 collaboration~\cite{ref:expt2}.  Dominance of the 
$qg$ component is evident over a large interval in $Q_T$, whereas the 
$q {\bar q}$ contribution begins to be felt at very large $Q_T$.

Data obtained by the Fermilab E772 collaboration~\cite{ref:expt1,ref:cfact} 
from an 800 GeV proton beam incident on a deuterium target are shown in 
Fig.~11 along with theoretical 
calculations.  The solid lines show the purely perturbative next-to-leading 
order expectation, and the dashed and dotted curves pertain to results 
obtained from all-orders soft-gluon resummation and its perturbative expansion 
at low $Q_T$.  

The leading-order ${\cal O}(\alpha_s)$ and next-to-leading order 
${\cal O}(\alpha_s^2)$ perturbative calculations of the $Q_T$ distribution 
necessarily fail for $Q_T << Q$, where the analytic form of the fixed-order 
cross section behaves as 
\begin{eqnarray}
d\sigma/dQ_T^2&\sim&\alpha_s/Q_T^2
\left\{1+a_1\alpha_s\ln^2(Q/Q_T)+a_2\alpha_s^2\ln^4(Q/Q_T)+...\right\}, 
\label{dyr}
\end{eqnarray}
In Eq.~(\ref{dyr}), $\alpha_s(\mu)$ is evaluated at the scale $\mu = Q$, 
and the $a_i$ are coefficients that do not depend on $Q_T$.    
The expansion parameter of the series in Eq.~(\ref{dyr}) is 
$\alpha_s\ln^2(Q/Q_T)$, not $\alpha_s$, and it can become large at small 
$Q_T$.  The logarithmic series, associated with soft-gluon emission, is a 
vestige of the singularities that cancel at each order in perturbation theory 
between the real gluon emission and virtual gluon exchange diagrams.  An 
elaborate procedure exists to sum the logarithmic series, known as 
resummation~\cite{ref:CSS,ref:DWS,ref:AEGM,ref:AK,ref:LY}.  We employ the 
formulation published by Arnold and Kauffman~\cite{ref:AK}, 
along with the phenomenological non-perturbative functions of 
Ladinsky and Yuan~\cite{ref:LY}.  We compare the resummed results 
with the purely perturbative results in order to try to establish the value 
above which we can be reasonably confident that fixed-order perturbation 
theory is adequate on its own.  

In Fig.~11, the dashed curves show the theoretical cross section obtained 
from resummation.  They are applicable only in the region $Q_T << Q$.  This 
restricted region of applicability arises because the only part of the 
fixed-order perturbative cross sections that is incorporated in the all-orders 
resummation formalism is the part that diverges as $Q_T^{-2}$ when 
$Q_T \rightarrow 0$.  The dotted curves in Fig.~11 show what is termed the 
``asymptotic" fixed-order contribution.  It is obtained from an expansion of 
the resummed cross section in a power series in $\alpha_s(Q)$.  Mathematically, 
it represents the divergent part of the fixed-order perturbative cross section 
asymptotically in the limit $Q_T \rightarrow 0$, and, as the coincidence of 
the dotted and solid curves shows in Fig.~11, this representation is exact at 
very small $Q_T$.  In our case, the dotted curves are the asymptotic answer 
through ${\cal O}(\alpha_s^2)$.

Whereas resummation is clearly needed at small $Q_T$, a fixed-order 
perturbative calculation, without resummation, should be adequate at large 
enough $Q_T$.  In addition to the singular parts incorporated in the 
resummation, fixed-order perturbative cross sections include remainders that 
are non-singular at small $Q_T$ but are important at large $Q_T$.  An issue 
addressed in the literature is how best to connect the region in which 
all-orders resummation is pertinent to the region in which the simpler 
fixed-order result is fine.  In efforts to restore the full content of 
fixed-order cross sections in the limit of large $Q_T$, proposals have been 
made whereby the non-singular parts are reintroduced through ``matching" 
prescriptions.  The results of one approach to matching are sketched in 
Fig.~11 as the dot-dashed lines.  The resummed result at very small $Q_T$ is 
joined smoothly to the purely perturbative fixed-order result at large $Q_T$.  

Our purpose is to establish where the data on massive lepton-pair production 
may be used to test fixed-order perturbative QCD and to provide new 
constraints on the gluon density.  The region of small $Q_T$ and the 
matching region of intermediate $Q_T$ are complicated by some level of 
phenomenological ambiguity.  Within the CSS resummation approach, 
phenomenological non-perturbative functions play a key role in fixing the shape 
of the $Q_T$ spectrum at very small $Q_T$, and matching methods in the 
intermediate region are hardly unique.  For the goals we have in mind, it would 
appear best to restrict attention to the region in $Q_T$ above the value at 
which the resummed result (dashed curves in Fig.~11) falls below the 
fixed-order perturbative expectation (solid curves).  A rough rule-of-thumb 
based on the calculations shown in Fig.~11 is $Q_T \geq Q/2$.  The published 
data~\cite{ref:expt1,ref:cfact} at fixed-target momenta shown in Figs.~11(a) 
and 11(b) do not extend into 
this domain.  Their reach is exhausted at just about the value in $Q_T$ at 
which the fixed-order calculation should be reliable.  

To establish the utility of massive lepton-pair production at fixed-target 
momenta for extraction of the gluon density, it would be valuable to obtain 
data in the traditional Drell-Yan region above $Q \simeq$ 4.0 GeV but whose 
reach in $Q_T$ extends into the fixed-order perturbative domain, as defined 
above.  The massive lepton-pair production cross section increases as $Q$ is 
reduced.  In Fig.~11(c) we show the predicted cross section as a function of 
$Q_T$ for 2.0 $< Q <$ 3.0 GeV.  The region $Q_T \geq$ 3.0 GeV is 
inviting.  It is above the region in which resummation need be considered, and 
as shown in Fig.~5(c), it is a region in which the $qg$ subprocess dominates 
the final cross section.  

In Fig.~12, we provide predictions of the the invariant inclusive cross 
section $Ed^3\sigma/d p^3$ as a function of $Q_T$ for 
$p + p\rightarrow \gamma^* + X$ at the LHC energy $\sqrt S =$ 14 TeV. The 
curves show the next-to-leading order perturbative predictions integrated over 
four different intervals of Q and averaged over the rapidity interval 
-1.0 $< y < $ 1.0.  In all four cases, the contribution from the 
$qg$ channel dominates at the level of 80 \% or greater over the range of 
$Q_T$ shown.  The LHC cross sections are about an order of magnitude greater 
than those at $\sqrt S =$ 1.8~TeV over the range of $Q_T$ shown.   


\section{Discussion and Conclusions}
\label{sec:5}

Prompt real photon production is a source of essential information on the 
{\it gluon} momentum density.  At lowest order in perturbation theory, the 
reaction is dominated at large values of the transverse momentum $p_T$ of the 
produced photon by the ``Compton" subprocess, $q + g \rightarrow \gamma + q$.  
This dominance by gluon initiated subprocesses is preserved at higher orders, 
indicating that the experimental 
inclusive cross section differential in $p_T$ may be used to determine the 
density of gluons in the initial hadrons.  There are notable similarities in 
the theoretical analyses of the massive lepton-pair production, the Drell-Yan 
process and prompt photon production.  At first-order in the strong coupling 
strength, $\alpha_s$, the Compton subprocess and the annihilation subprocess 
$q +\bar{q} \rightarrow \gamma + g$ supply the transverse momentum of the 
photons in both cases.  The parallels are maintained in next-to-leading order, 
as summarized in Sec.~II.

In this paper we focus on the $Q_T$ distribution for $h_1 + h_2 \rightarrow 
\gamma^* + X$.  We present and discuss calculations carried out in 
next-to-leading order QCD at both fixed target and collider energies.  We 
show that the differential cross section in the region $Q_T \geq Q/2$ is 
dominated by subprocesses initiated by incident gluons.  Dominance of the 
$qg$ contribution in the massive lepton-pair case is as strong if not stronger 
than it is in the prompt photon case.  Massive lepton-pair 
differential cross sections are therefore an additional useful source of 
constraints on the gluon density.  We compare calculations with data and 
provide predictions for the differential cross section as a function of $Q_T$ 
in proton-antiproton reactions at a center-of-mass energy of 1.8~TeV, 
in proton-nucleon reactions at laboratory momentum 800 GeV, and in 
proton-proton reactions at LHC energies.  

As long $Q_T$ is large, the perturbative requirement of small $\alpha_s(Q_T)$ 
can be satisfied without a large value of $Q$.  We therefore explore and 
advocate the potential advantages of studies of $d^2\sigma/dQ dQ_T$ as a 
function of $Q_T$ for modest values of $Q$,  
$Q \sim 2$GeV, below the range of the traditional Drell-Yan region. 
   
As our calculations demonstrate, the $Q_T$ distribution of massive lepton 
pair production offers a valuable additional method for direct measurement of 
the gluon momentum distribution.  This method makes good use of a significant 
data base at collider and fixed target energies.  The method is similar in 
principle to the approach based on prompt photon production, but it avoids 
the experimental and theoretical complications of photon isolation that beset 
studies of prompt photon production.  No isolation would seem necessary in the 
case of massive virtual photon 
production (and subsequent decay into a pair of muons) in typical collider or 
fixed target experiments.  Muons are observed only after they have penetrated a 
substantial hadron absorber.  Thus, hadrons within a typical cone about 
the direction of the $\gamma^*$ will have been stopped, and the massive 
lepton-pair signal will be entirely inclusive.  For this reason, we claim 
that the cross section for massive lepton-pair production at large $Q_T$ is 
an {\it advantageous} source of information on the gluon density.  

All-orders resummation is known to be important for the description of the 
$Q_T$ distribution of massive lepton-pair production at small and modest values 
of $q_T$.  In this paper, we compare the resummed results 
with the purely perturbative results in order to try to establish the value 
above which we can be reasonably confident that fixed-order perturbation 
theory is adequate on its own.  For the goals we have in mind, it would 
appear best to restrict attention to the region in $Q_T$ above the value at 
which the resummed result falls below the fixed-order perturbative expectation. 
A rough rule-of-thumb based on our calculations is $Q_T \geq Q/2$.  
The published data~\cite{ref:expt1} at fixed-target momenta do not 
extend into this domain.  Their reach is exhausted at just about the value in 
$Q_T$ at which the fixed-order calculation should be reliable.  
To establish the utility of massive lepton-pair production at fixed-target 
momenta for extraction of the gluon density, it would be valuable to obtain 
data in the traditional Drell-Yan region above $Q \simeq$ 4.0 GeV but whose 
reach in $Q_T$ extends into the fixed-order perturbative domain.  The
experimental rates at the $p {\bar p}$ collider energy of $\sqrt S =$ 1.8 TeV 
should be adequate over a wide range in $Q_T$.  


\section*{Acknowledgments}

Work in the High 
Energy Physics Division at Argonne National Laboratory is supported by 
the U.S. Department of Energy, Division of High Energy Physics, 
Contract W-31-109-ENG-38.  We have benefitted from communications with 
C. Brown, G. T. Garvey, and J. Moss concerning their data from FNAL 
fixed-target experiments and are grateful to P. Arnold, R. Kauffman, and 
M.~H.~Reno for providing copies of their numerical codes.  



\begin{figure}
 \begin{center}
  {\unitlength1cm
  \begin{picture}(15,17)
   \epsfig{file=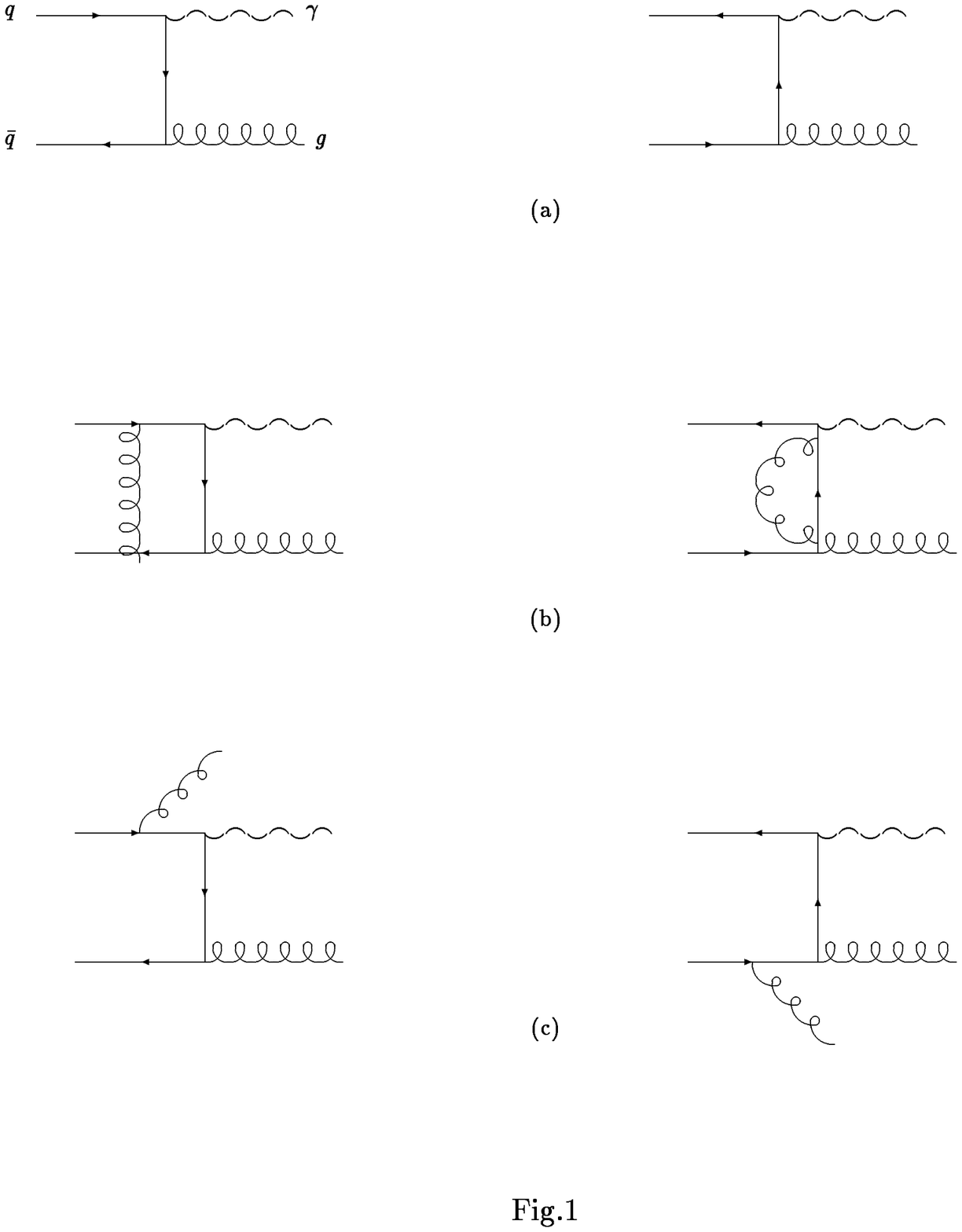,bbllx=95pt,bblly=155pt,bburx=540pt,bbury=645pt,%
           height=17cm,clip=}
  \end{picture}}
 \end{center}
\caption{(a)Lowest-order Feynman diagrams for the direct process 
$q + \bar{q} \rightarrow \gamma + g$. (b) Examples of virtual gluon loop 
diagrams.  (c) Examples of next-to-leading order three-body final-state 
diagrams.}
\label{fig1}
\end{figure}

\begin{figure}
 \begin{center}
  {\unitlength1cm
  \begin{picture}(15,20)
   \epsfig{file=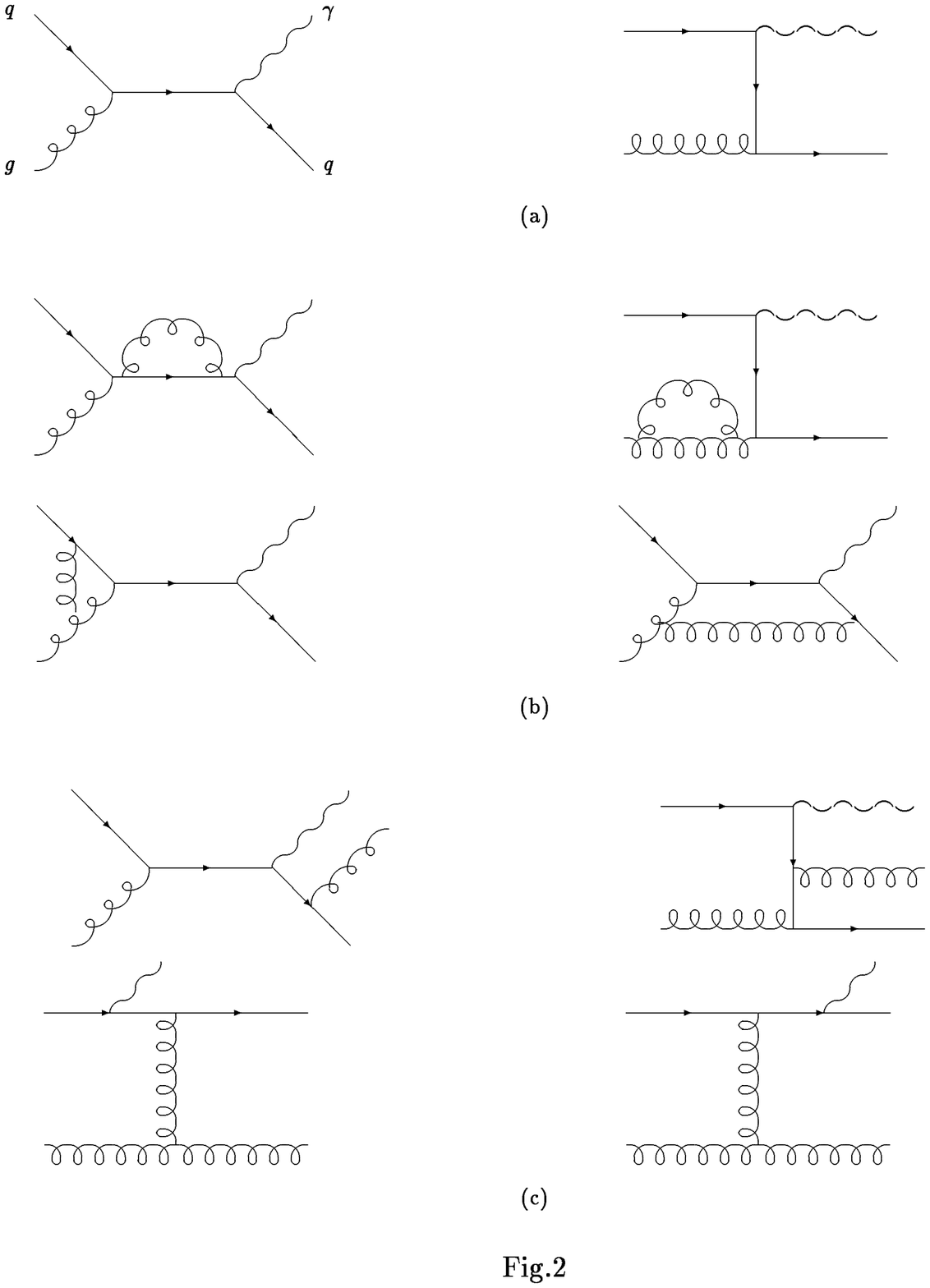,bbllx=85pt,bblly=60pt,bburx=540pt,bbury=655pt,%
           height=20cm,clip=}
  \end{picture}}
 \end{center}
\caption{As in Fig.~1, but for the subprocesses initiated by the $q + g$ 
initial state.}
\label{fig2}
\end{figure}

\begin{figure}
 \begin{center}
  {\unitlength1cm
  \begin{picture}(15,10)
   \epsfig{file=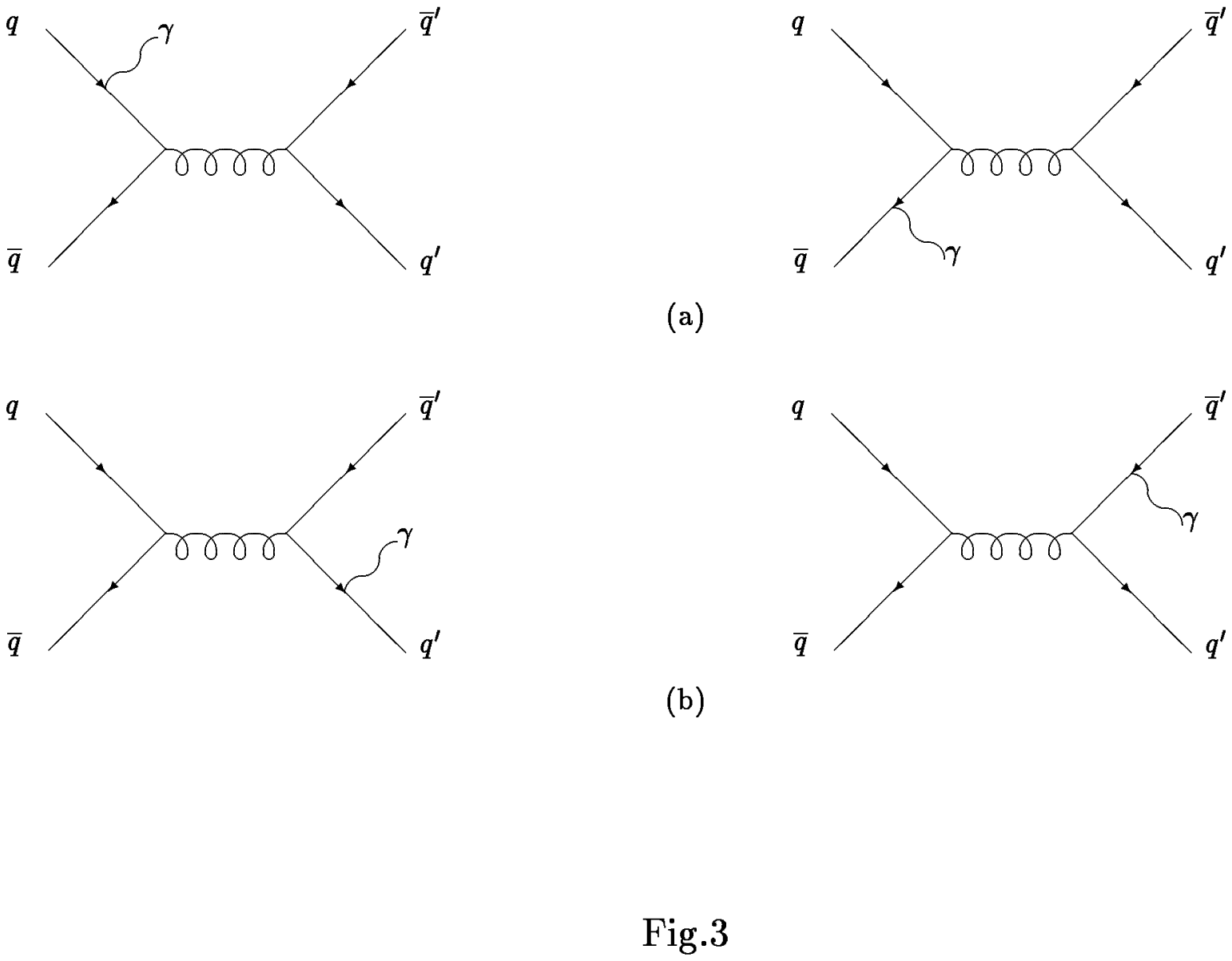,bbllx=100pt,bblly=385pt,bburx=545pt,bbury=650pt,%
           width=15cm,clip=}
  \end{picture}}
 \end{center}
\caption{Diagrams that illustrate situations in which the photon becomes 
collinear to a quark in (a) the initial state and (b) the final state.}
\label{fig3}
\end{figure}

\begin{figure}
 \begin{center}
  {\unitlength1cm
  \begin{picture}(12,18)
   \epsfig{file=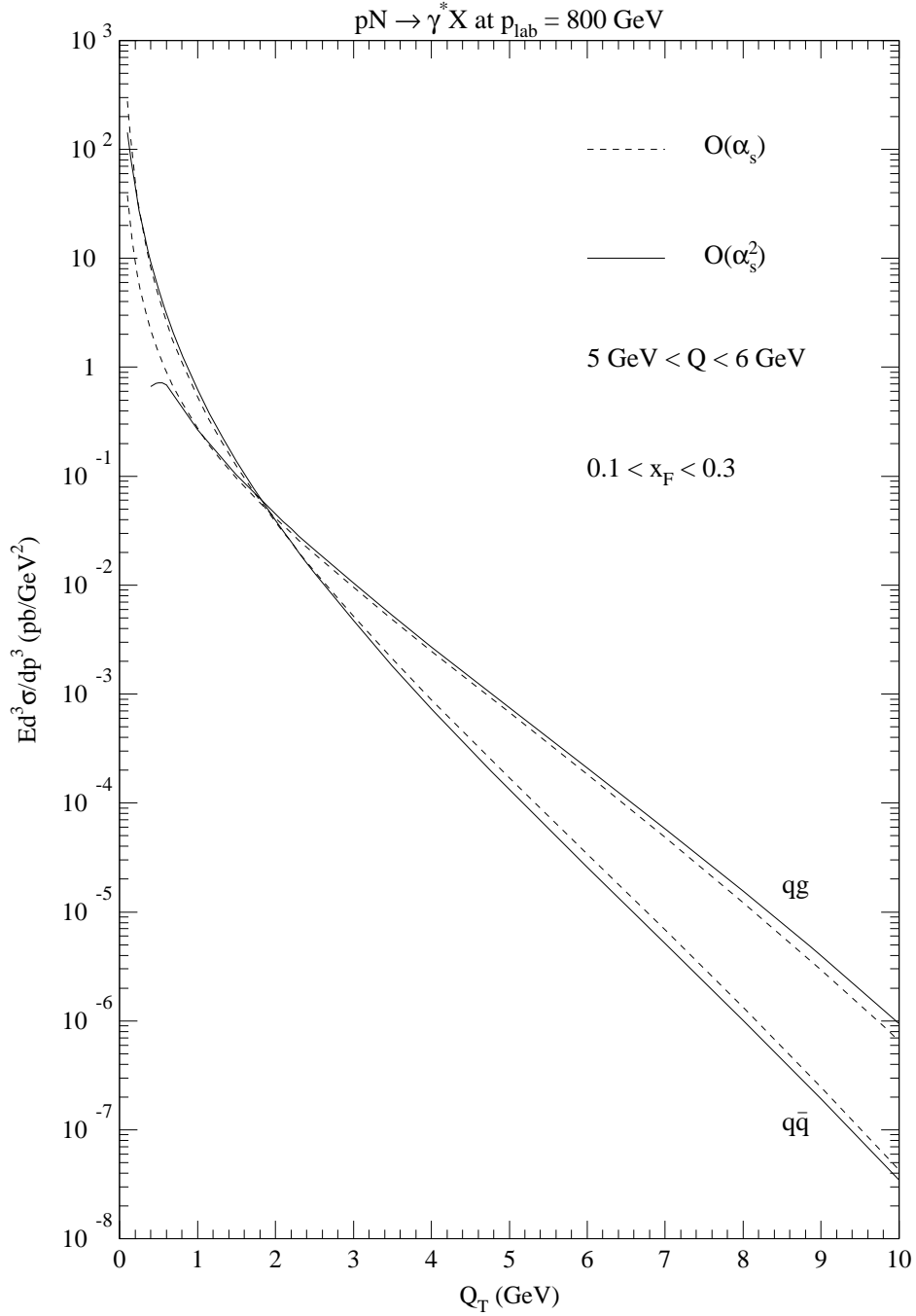,bbllx=55pt,bblly=100pt,bburx=495pt,bbury=725pt,%
           height=18cm}
  \end{picture}}
 \end{center}
\caption{Lowest order ${\cal O}(\alpha_s)$ (dashed lines) and next-to-leading 
order ${\cal O}(\alpha_s^2)$ (solid lines) perturbative calculations of the 
invariant inclusive cross section $Ed^3\sigma/dp^3$ as a function of $Q_T$ for 
$p N \rightarrow \gamma^* X$ at $p_{\rm lab}$ = 800 GeV, in the 
$\overline{\rm MS}$ scheme.  Contributions from the $qg$ and $q \bar{q}$ 
channels are shown separately.  The results are integrated over the scaled 
longitudinal momentum interval 0.1 $<x_F<$ 0.3 and over the interval 
5.0 $<Q<$ 6.0 GeV, and they are divided by the bin width in $x_F$.}
\label{fig4}
\end{figure}

\begin{figure}
 \begin{center}
  {\unitlength1cm
  \begin{picture}(12,18)
   \epsfig{file=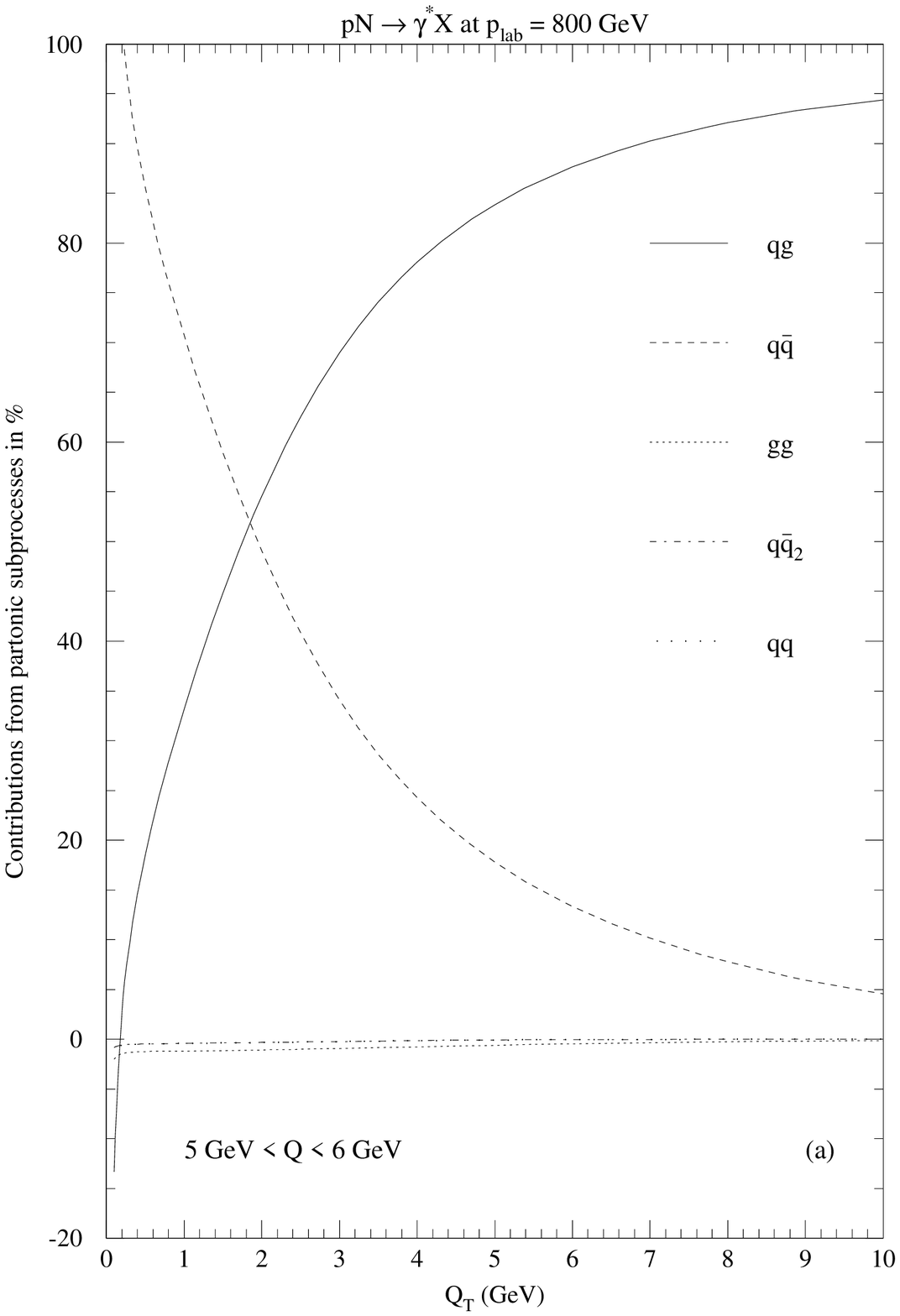,bbllx=55pt,bblly=100pt,bburx=495pt,bbury=725pt,%
           height=18cm}
  \end{picture}}
 \end{center}
 \begin{center}
  {\unitlength1cm
  \begin{picture}(12,18)
   \epsfig{file=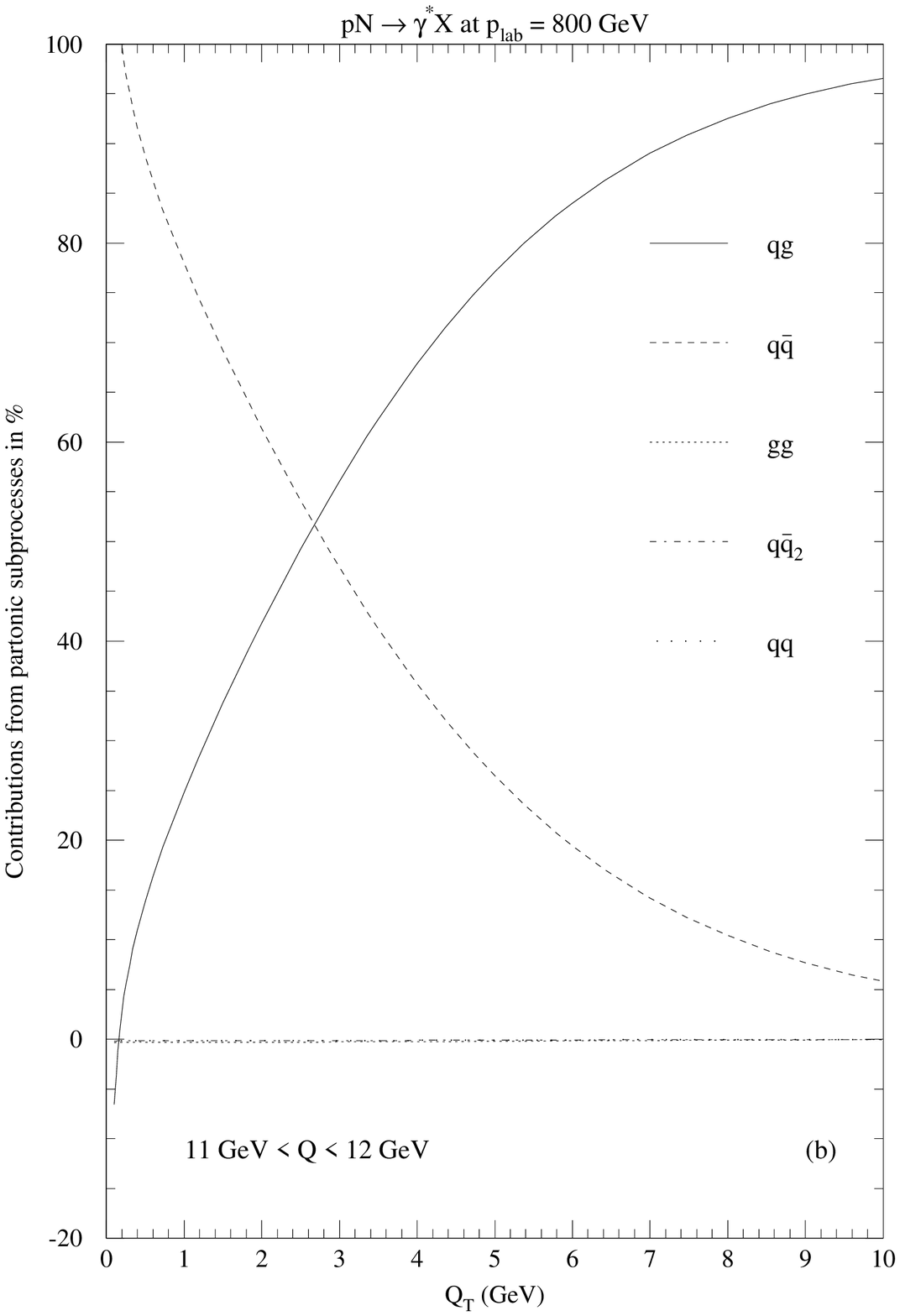,bbllx=55pt,bblly=100pt,bburx=495pt,bbury=725pt,%
           height=18cm}
  \end{picture}}
 \end{center}
 \begin{center}
  {\unitlength1cm
  \begin{picture}(12,18)
   \epsfig{file=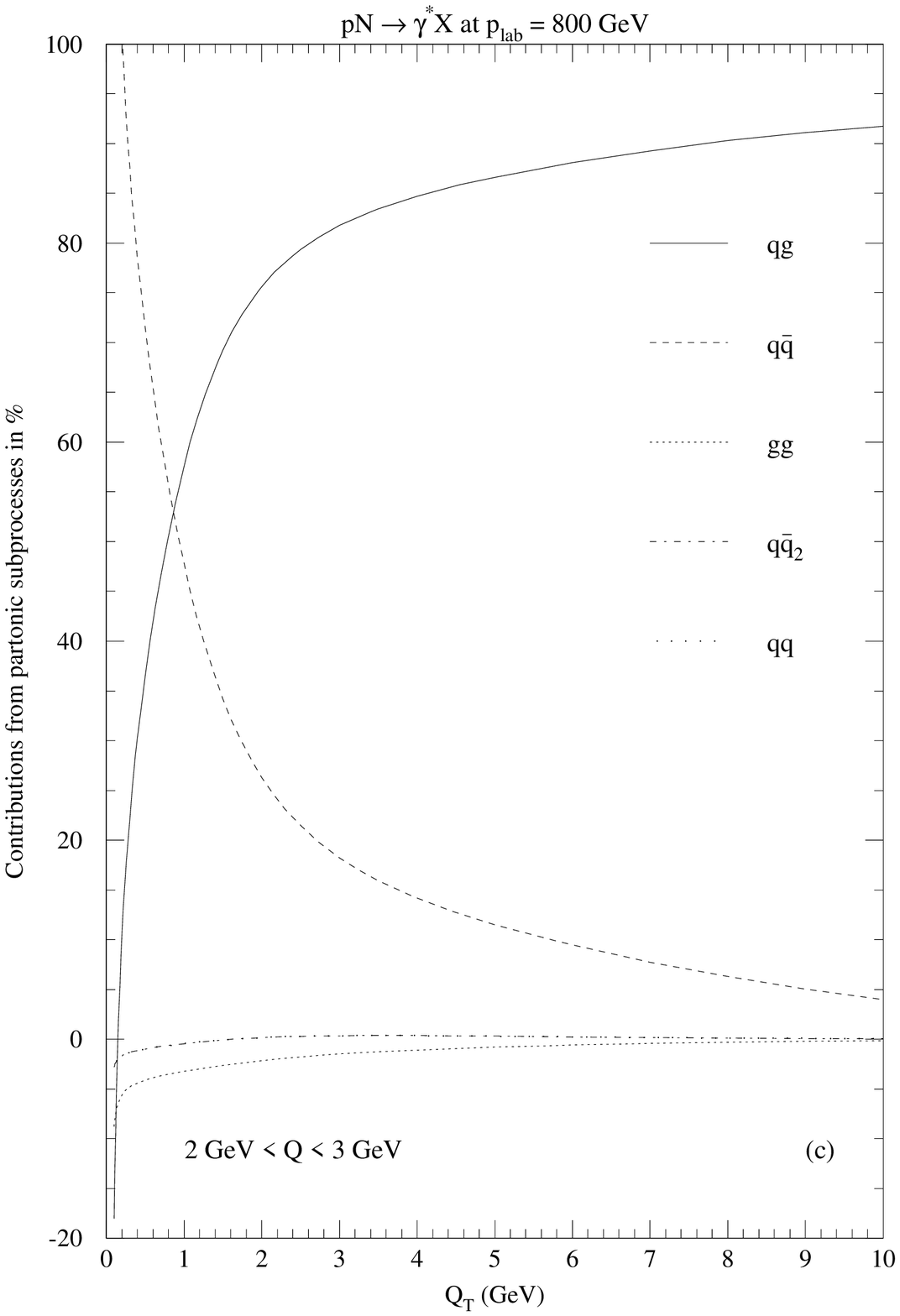,bbllx=55pt,bblly=100pt,bburx=495pt,bbury=725pt,%
           height=18cm}
  \end{picture}}
 \end{center}
\caption{Contributions from the various partonic subprocesses to the invariant 
inclusive cross section $Ed^3\sigma/dp^3$ as a function of $Q_T$ for 
$p N \rightarrow \gamma^* X$ at $p_{\rm lab}$ =
800 GeV.  The cross section is integrated over the scaled longitudinal momentum 
interval 0.1 $<x_F<$ 0.3 and over the intervals (a) 5.0 $<Q<$ 6.0 GeV, (b) 
11.0 $<Q<$ 12.0 GeV, and (c) 2.0 $<Q<$ 3.0 GeV, and divided by the bin width 
in $x_F$. The contributions are labeled by $qg$ (solid), $q \bar{q}$ (dashed), 
$gg$ (dotted), $q \bar{q}_2$ non-factorizable parts (dot-dashed), and $qq$ 
(wide dots).  The $qq$ and $q \bar{q}_2$ results nearly coincide.}
\label{fig5}
\end{figure}

\begin{figure}
 \begin{center}
  {\unitlength1cm
  \begin{picture}(12,18)
   \epsfig{file=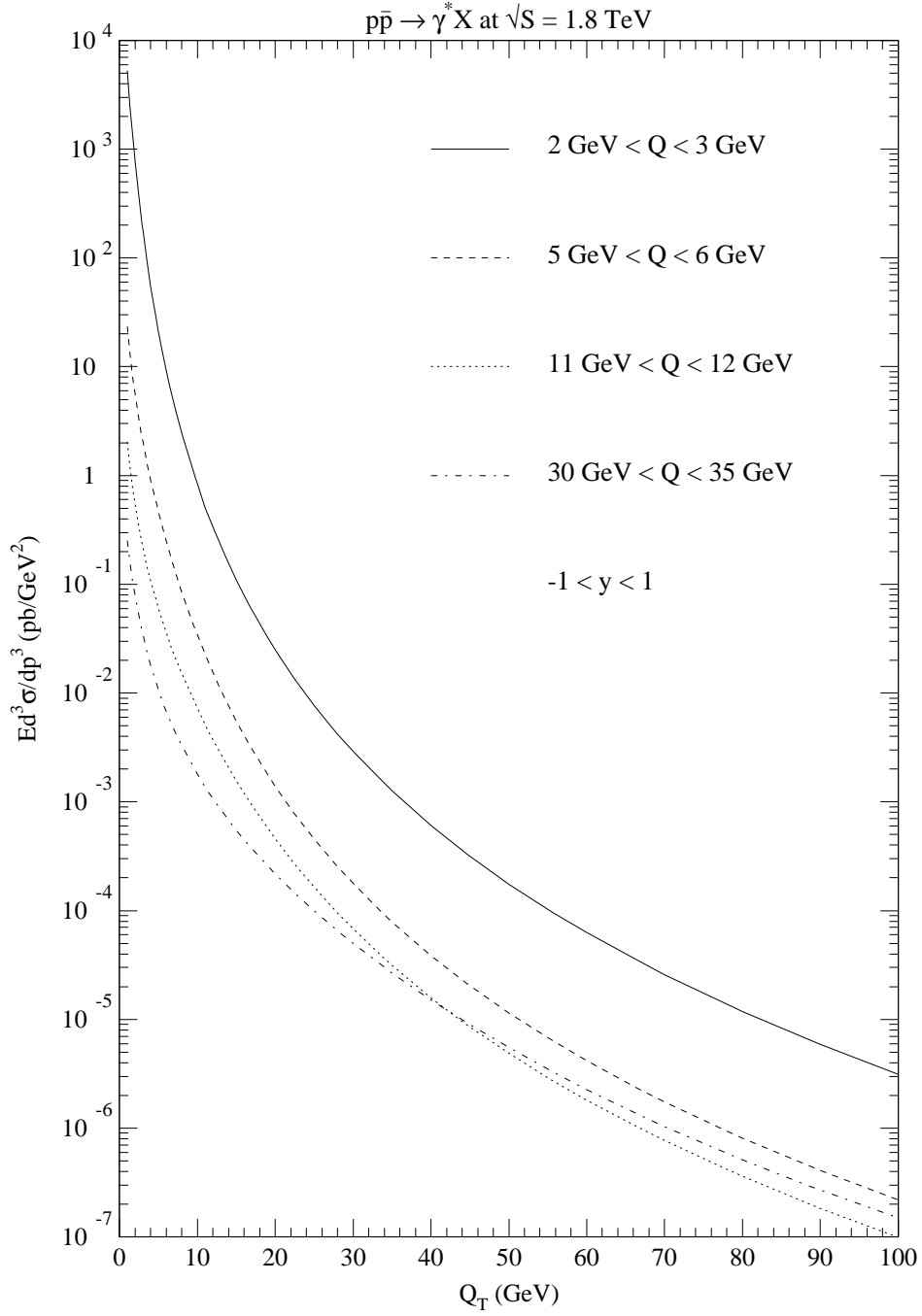,bbllx=55pt,bblly=100pt,bburx=495pt,bbury=725pt,%
           height=18cm}
  \end{picture}}
 \end{center}
\caption{Differential cross sections $Ed^3\sigma/dp^3$ as a function of $Q_T$ 
for $p \bar{p} \rightarrow \gamma^* X$ at $\protect\sqrt{S}=1.8$ TeV 
integrated over the rapidity interval -1.0 $< y <$ 1.0 and over the intervals 
2.0 $<Q<$ 3.0 GeV (solid), 5.0 $<Q<$ 6.0 GeV (dashed), 11.0 $<Q<$ 12.0 GeV 
(dotted), and 30.0 $<Q<$ 35.0 GeV (dot-dashed). The results of the integration 
have been divided by the bin width in $y$.}   
\label{fig6}
\end{figure}

\begin{figure}
 \begin{center}
  {\unitlength1cm
  \begin{picture}(12,18)
   \epsfig{file=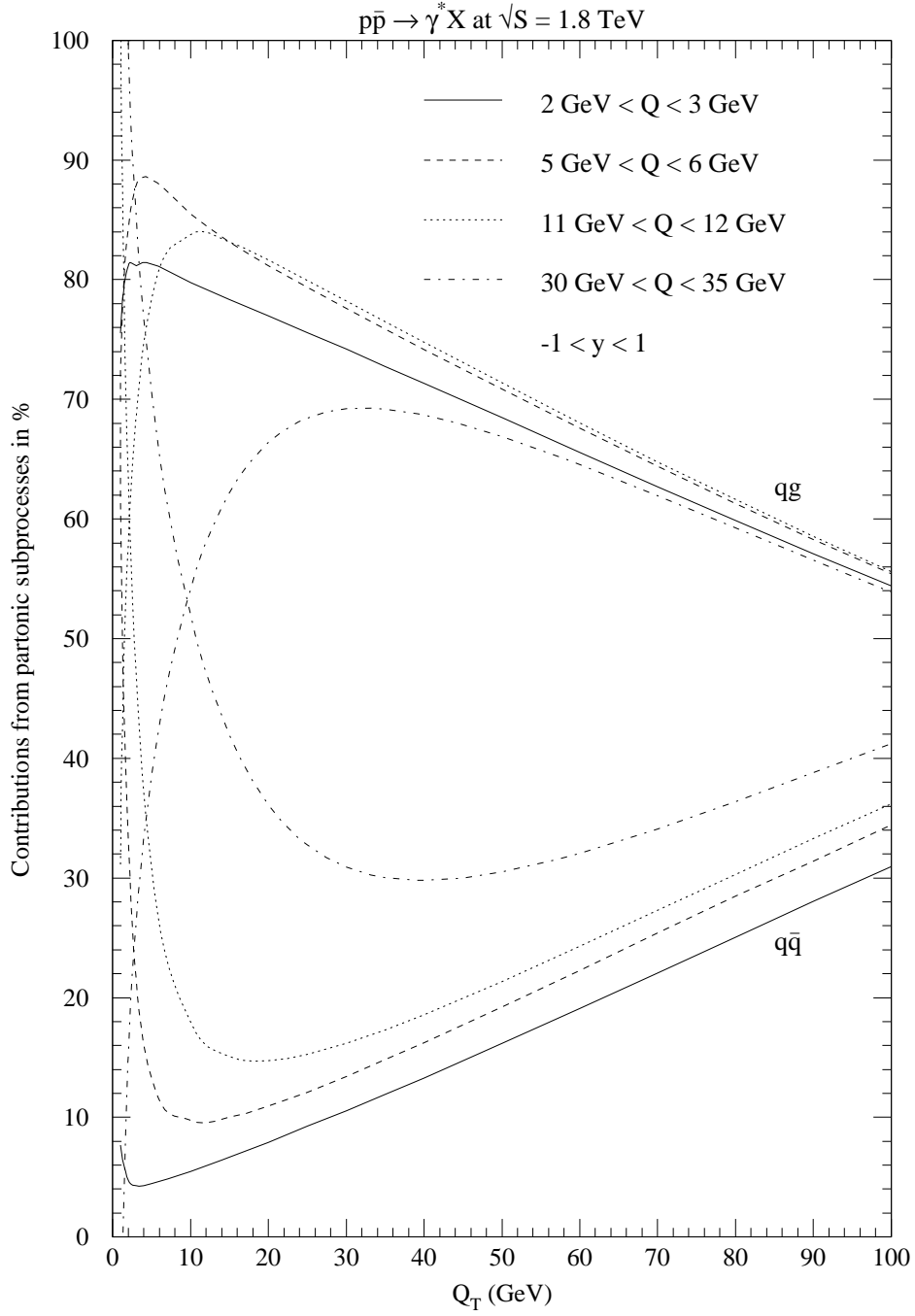,bbllx=55pt,bblly=100pt,bburx=495pt,bbury=725pt,%
           height=18cm}
  \end{picture}}
 \end{center}
\caption{Contributions from the partonic subprocesses $qg$ and $q \bar{q}$ 
to the invariant 
inclusive cross section $Ed^3\sigma/dp^3$ as a function of $Q_T$ for 
$p {\bar p}\rightarrow \gamma^* X$ at $\protect\sqrt{S}$ =
1.8 TeV.  The cross section is integrated over the rapidity  
interval -1.0 $<y<$ 1.0 and over the intervals 2.0 $<Q<$ 3.0 GeV (solid),  
5.0 $<Q<$ 6.0 GeV (dashed), 11.0 $<Q<$ 12.0 GeV (dotted), and 
30.0 $<Q<$ 35.0 GeV (dot-dashed). The results of the integration have been 
divided by the bin width in $y$.} 
\label{fig7}
\end{figure}

\begin{figure}
 \begin{center}
  {\unitlength1cm
  \begin{picture}(12,18)
   \epsfig{file=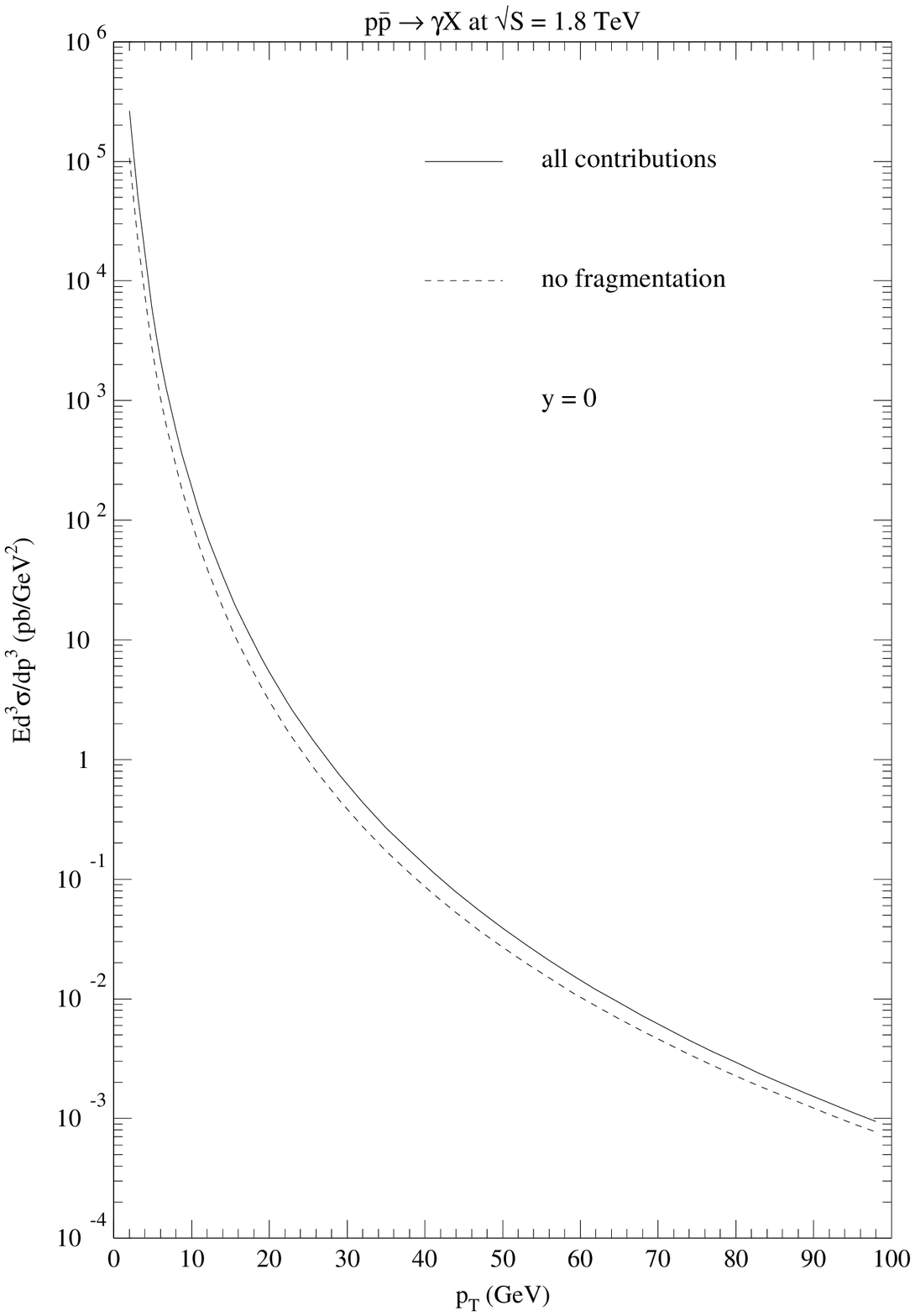,bbllx=55pt,bblly=100pt,bburx=495pt,bbury=725pt,%
           height=18cm}
  \end{picture}}
 \end{center}
\caption{Differential cross sections $Ed^3\sigma/dp^3$ as a function of $p_T$  
for real photon production $p \bar{p} \rightarrow \gamma X$ at 
$\protect\sqrt{S}=1.8$ TeV in the $\overline{\rm MS}$-scheme for two 
cases: no fragmentation terms included (dashed) and the inclusive case 
with full fragmentation included (solid). The rapidity $y =$ 0.}
\label{fig8}
\end{figure}

\begin{figure}
 \begin{center}
  {\unitlength1cm
  \begin{picture}(12,18)
   \epsfig{file=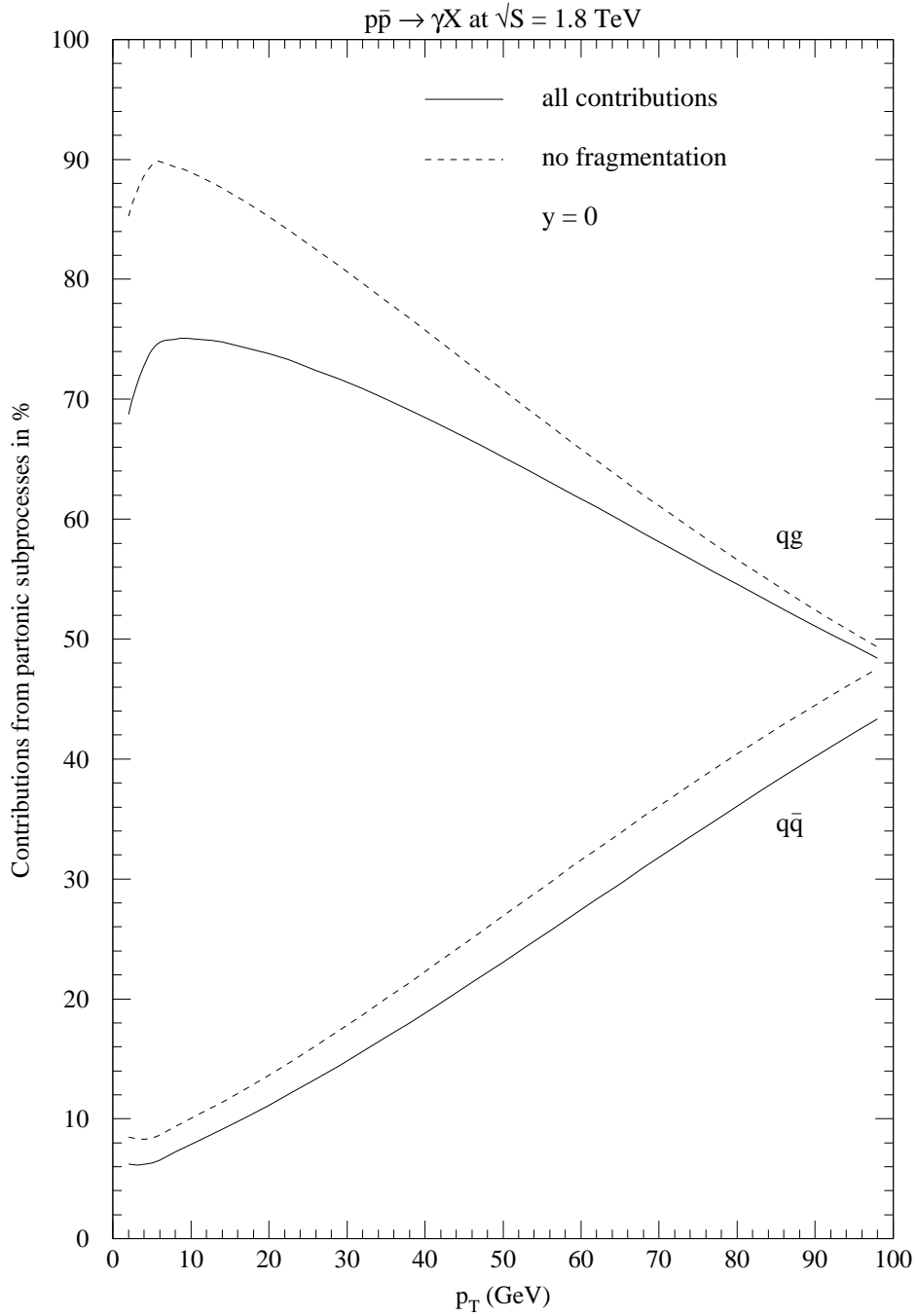,bbllx=55pt,bblly=100pt,bburx=495pt,bbury=725pt,%
           height=18cm}
  \end{picture}}
 \end{center}
\caption{Separate contributions from the $q\bar{q}$ and the $qg$ channels
to the differential cross sections $Ed^3\sigma/dp^3$ as a function of $p_T$  
for real photon production $p \bar{p} \rightarrow \gamma X$ at 
$\protect\sqrt{S}=1.8$ TeV in the $\overline{\rm MS}$-scheme for two 
cases: no fragmentation terms included (dashed lines) and the inclusive case 
with full fragmentation included (solid lines). The rapidity $y =$ 0.}
\label{fig9}
\end{figure}

\begin{figure}
 \begin{center}
  {\unitlength1cm
  \begin{picture}(12,18)
   \epsfig{file=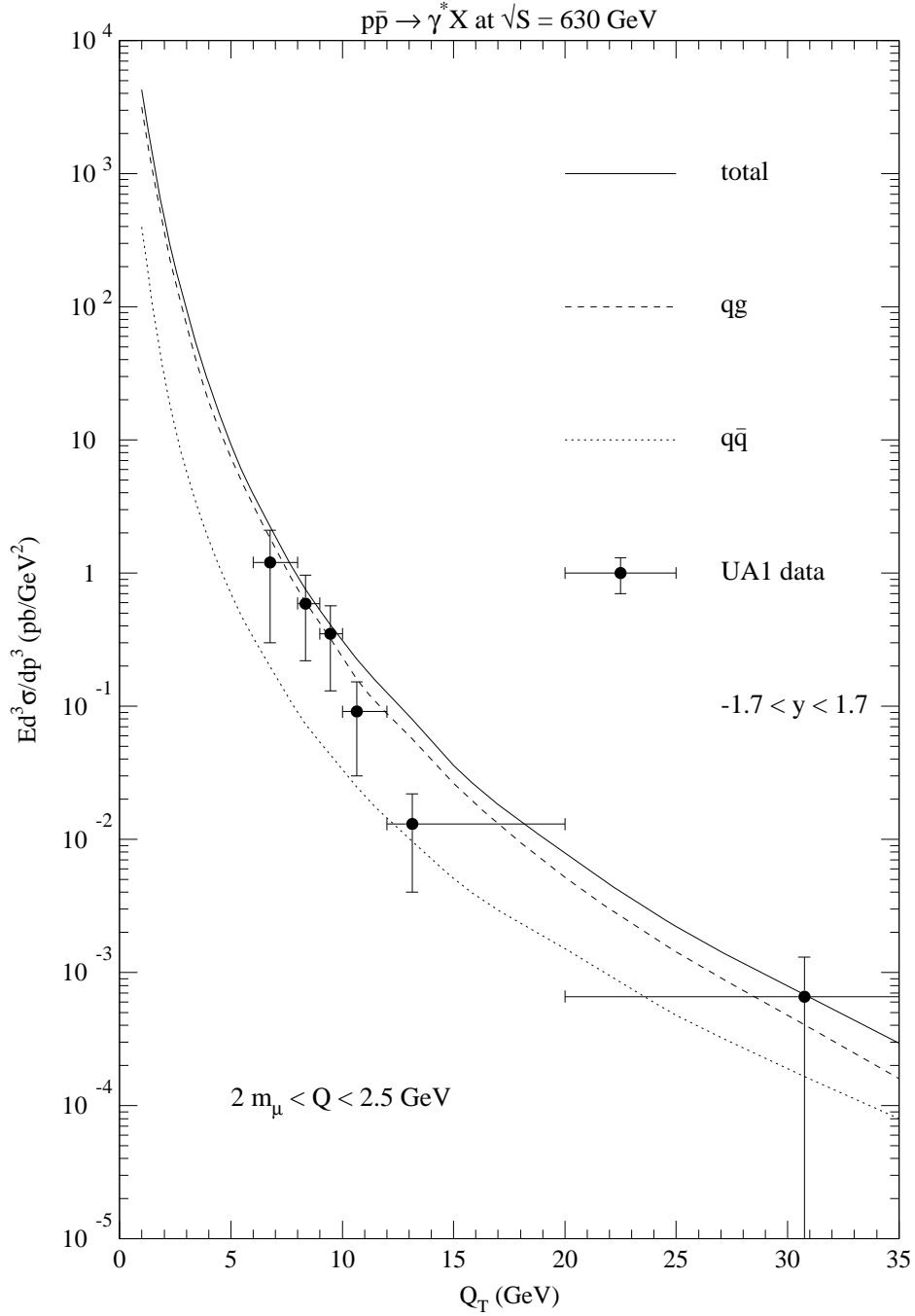,bbllx=55pt,bblly=100pt,bburx=495pt,bbury=725pt,%
           height=18cm}
  \end{picture}}
 \end{center}
\caption{Invariant inclusive cross section $Ed^3\sigma/d p^3$ as a function of 
$Q_T$ for $p \bar{p} \rightarrow \gamma^* +X$ at $\protect\sqrt{S}=$ 
630 GeV, with $2 m_{\mu} \leq Q \leq$ 2.5 GeV, and averaged over the rapidity 
interval -1.7 $< y <$1.7.  The next-to-leading order perturbative cross 
section (solid) is shown along with its two major components, the $qg$ 
(dashed) and $q \bar{q}$ (dotted) contributions. The data are from 
the CERN UA1 collaboration.}    
\label{fig10}
\end{figure}

\begin{figure}
 \begin{center}
  {\unitlength1cm
  \begin{picture}(12,18)
   \epsfig{file=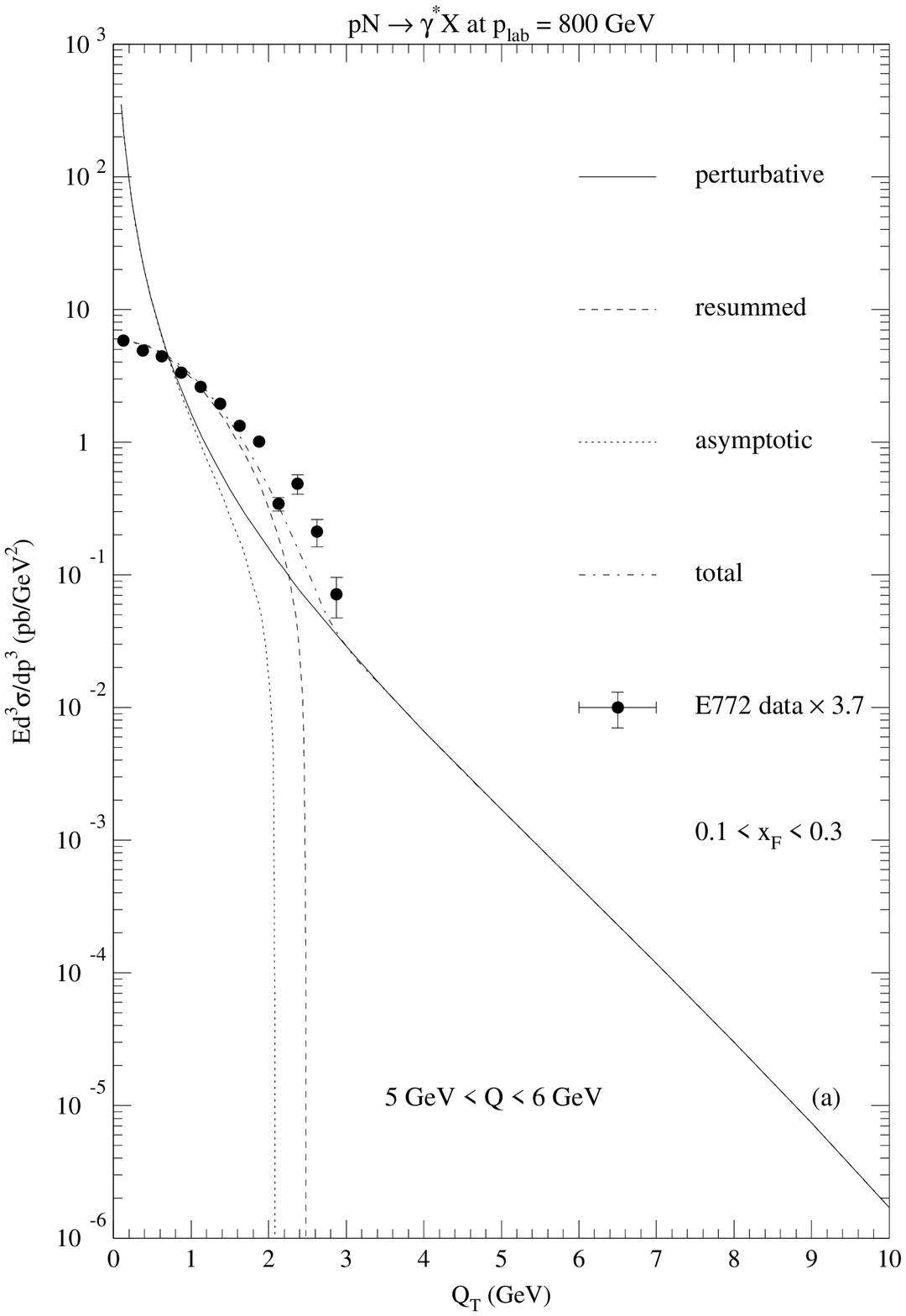,bbllx=55pt,bblly=100pt,bburx=495pt,bbury=725pt,%
           height=18cm}
  \end{picture}}
 \end{center}
 \begin{center}
  {\unitlength1cm
  \begin{picture}(12,18)
   \epsfig{file=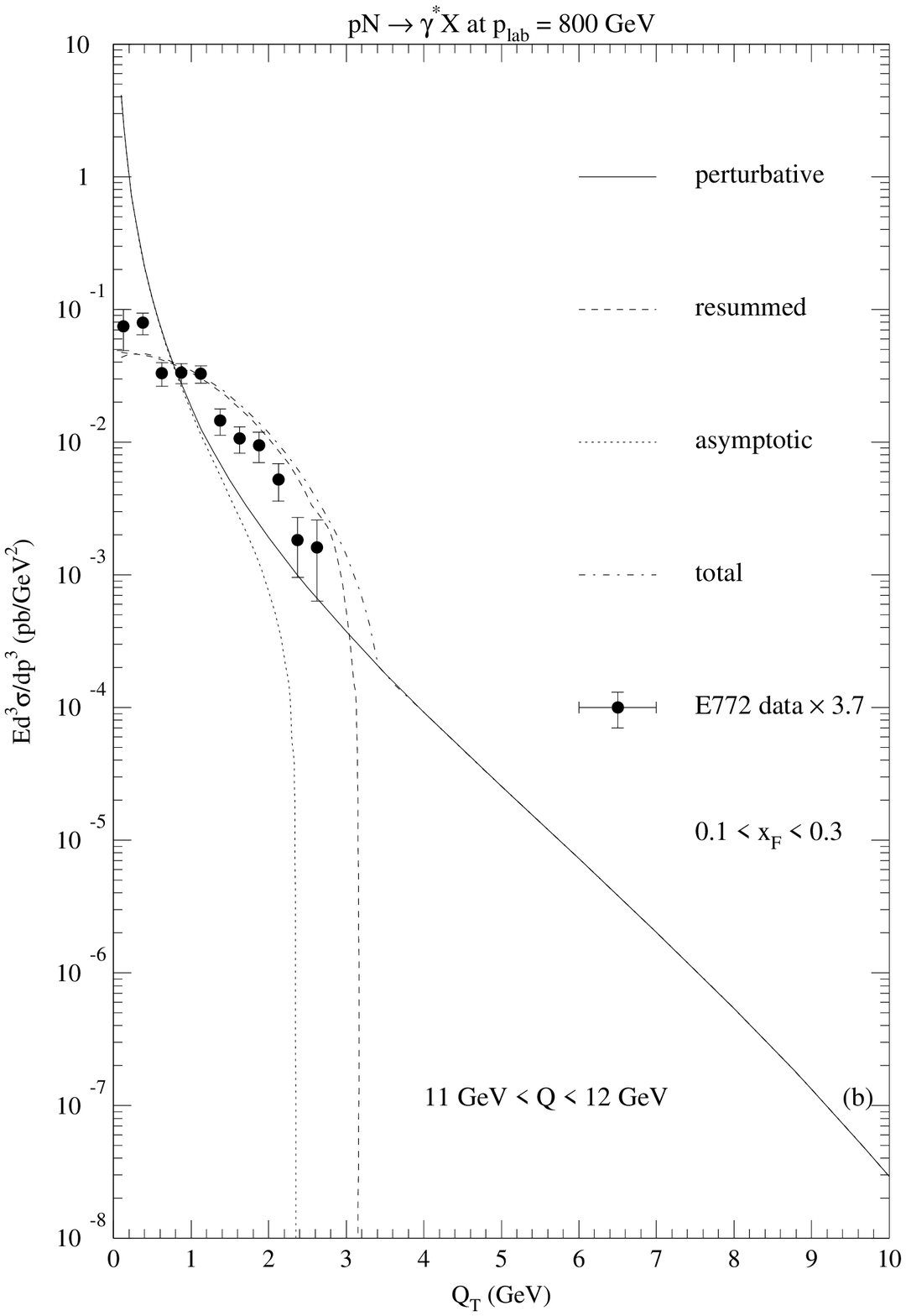,bbllx=55pt,bblly=100pt,bburx=495pt,bbury=725pt,%
           height=18cm}
  \end{picture}}
 \end{center}
 \begin{center}
  {\unitlength1cm
  \begin{picture}(12,18)
   \epsfig{file=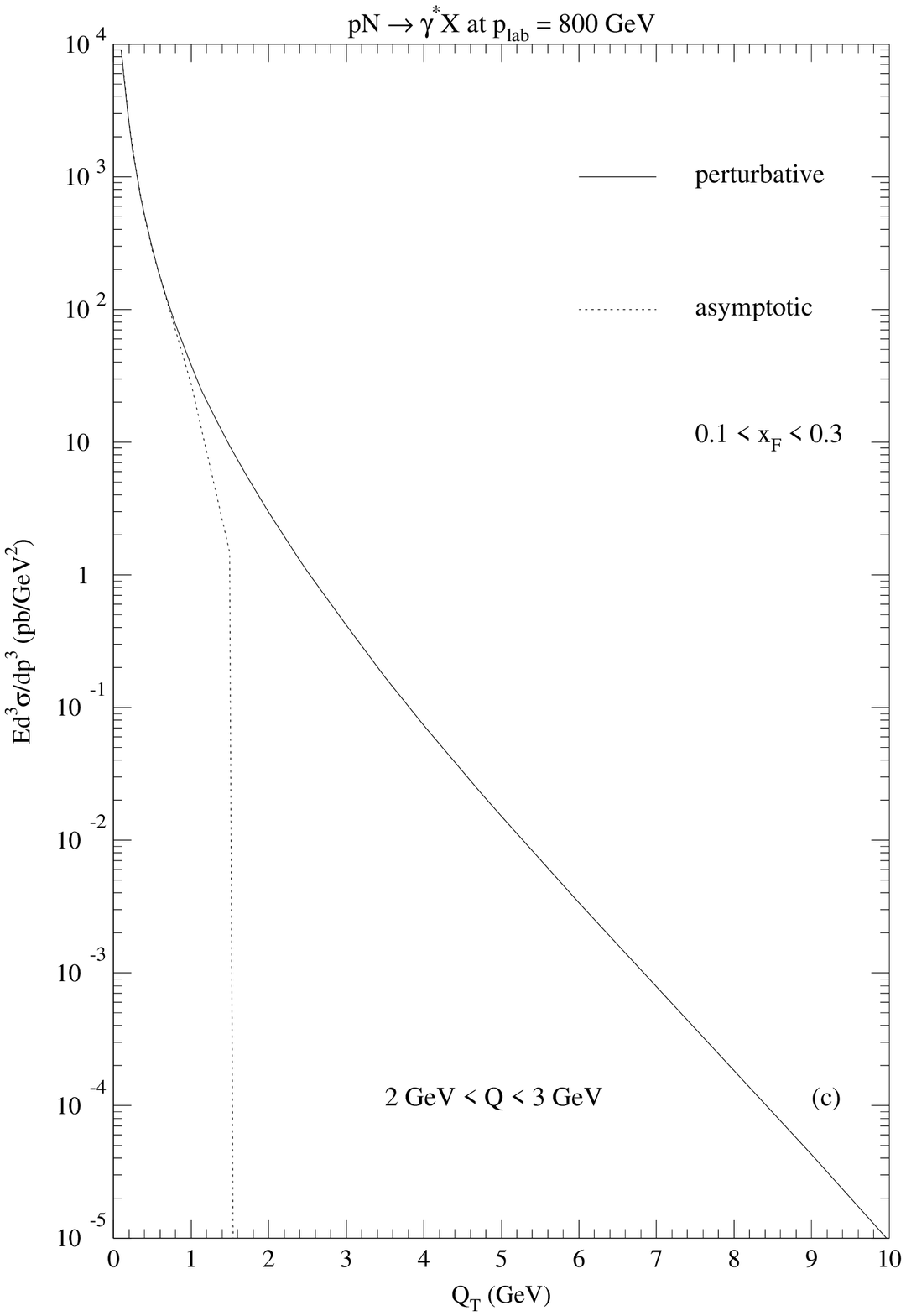,bbllx=55pt,bblly=100pt,bburx=495pt,bbury=725pt,%
           height=18cm}
  \end{picture}}
 \end{center}
\caption{Invariant inclusive cross section $Ed^3\sigma/d p^3$ as a function of 
$Q_T$ for $p N \rightarrow \gamma^* +X$ at $p_{\rm lab}=$ 800 GeV, 
averaged over the scaled longitudinal momentum interval 0.1 $< x_F <$ 0.3.  
The next-to-leading order perturbative cross section (solid) is shown along 
with the all-orders resummed expectation (dashed), the asymptotic result 
(dotted), and a matched expression (dot-dashed).  The data are from 
the Fermilab E772 collaboration.  (a) 5.0 $<Q<$ 6.0 GeV, (b) 
11.0 $<Q<$ 12.0 GeV, and (c) 2.0 $<Q<$ 3.0 GeV.} 
\label{fig11}
\end{figure}

\begin{figure}
 \begin{center}
  {\unitlength1cm
  \begin{picture}(12,18)
   \epsfig{file=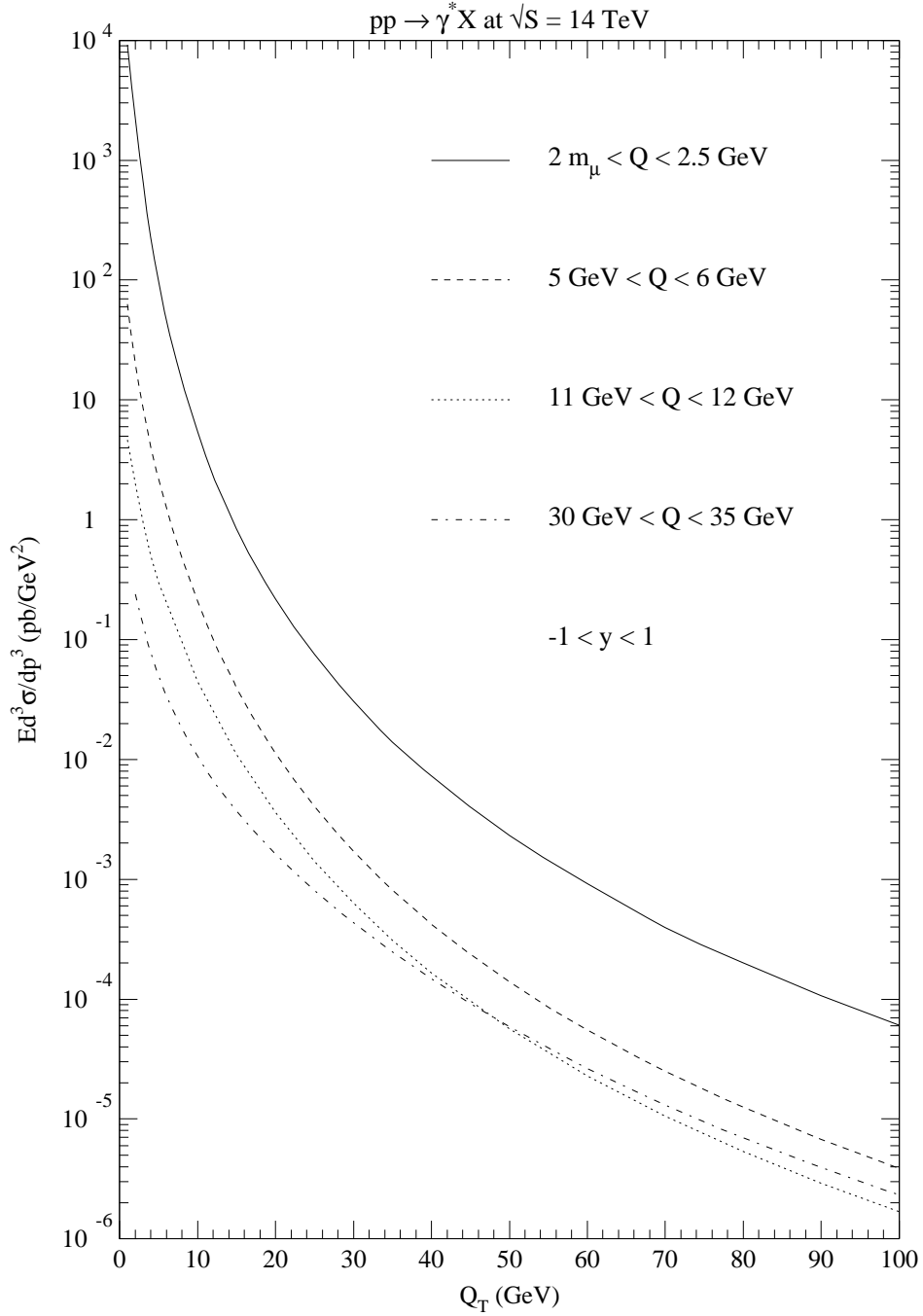,bbllx=55pt,bblly=100pt,bburx=495pt,bbury=725pt,%
           height=18cm}
  \end{picture}}
 \end{center}
\caption{Invariant inclusive cross section $Ed^3\sigma/d p^3$ as a function of 
$Q_T$ for $p p \rightarrow \gamma^* +X$ at $\protect\sqrt{S}=$ 
14 TeV averaged over the rapidity 
interval -1.0 $< y <$1.0.  The four curves show the next-to-leading order 
perturbative predictions integrated over 4 different intervals of Q: 
$2m_{\mu}<Q<$ 2.5 GeV, 5.0 $<Q<$ 6.0 GeV, 11.0 $<Q<$ 12.0 GeV, and 
30.0 $<Q<$ 35.0 GeV.}    
\label{fig12}
\end{figure}


\end{document}